\documentclass{aa}

\usepackage[utf8]{inputenc}
\usepackage[varg]{txfonts}
\usepackage{graphicx}
\usepackage{empheq}
\usepackage{natbib}
\bibpunct{(}{)}{;}{a}{}{,} 

\begin{document}

\title{Unbeamed tidal disruption events at hard X-rays}

\author{K. Hryniewicz\inst{1}\and R. Walter\inst{1}}
\institute{ISDC Data Centre for Astrophysics, Observatoire de Gen\`eve, Universit\'e de Gen\`eve, Chemin d’Ecogia 16, 1290 Versoix,  Switzerland\newline
\email{krhr@camk.edu.pl}}

\date{Received 26.05.2015; accepted 22.10.2015}

 
\abstract
   {Owing to their thermal emission, tidal disruption events (TDEs) were regularly detected  in the soft X-rays and sometimes in the optical. Only a few TDEs have been detected at hard X-rays: two are high redshift beamed events, one of which occurred at the core of a nearby galaxy, and the most recent one is of a different nature, involving a compact object in the Milky Way.}
   {The aims of this work are to obtain a first sample of hard X-ray-selected unbeamed TDEs, to determine their frequency and to probe whether TDEs  usually or exceptionally emit at hard X-ray energies.}
   {We performed extensive searches for hard X-ray flares at  positions in over 53\,000 galaxies, up to a distance of 100 Mpc in the {\it Swift} BAT archive. Light curves were extracted and parametrized. The quiescent hard X-ray emission was used to exclude persistently active galactic nuclei. Significant flares from non-active galaxies were derived and checked for possible contamination.}
   {We found a sample of nine TDE candidates, which translates into  a rate of $2 \times 10^{-5}\,{\rm galaxy}^{-1}\,{\rm yr}^{-1}$ above the BAT detection limit. This rate is
consistent with those observed by {\it XMM-Newton} at soft X-rays and in the optical from SDSS observations, and is as expected from simulations. We conclude that hard X-ray emission should be ubiquitous in un-beamed TDEs and that electrons should be accelerated in their accretion flow.}
   {}

\authorrunning{K. Hryniewicz \& R. Walter}
\titlerunning{Unbeamed TDEs at hard X-rays}

\maketitle

\section{Introduction}

Tidal disruption events (TDEs) occur when the gravitational potential of a compact object is large enough to tear apart an approaching star or planet. The fate of the disrupted object depends on whether it approaches the tidal radius $r_t=(2\,M_{\rm BH}/M_{*})^{1/3}\, R_{*}$, where $M_{\rm BH}$ is the compact object mass and $M_{*}$ and $R_{*}$ are the mass and radius of the disrupting object. Part of the debris resulting from a disruption will fall on the compact object at a rate which depends on the disrupted object's impact parameter and internal structure (binding energy). The accreted debris will heat up, might form a disc, accelerate particles in shocks, and possibly launch a jet. The heated material will shine in the optical and soft X-rays, accelerated particles may radiate in the hard X-rays through inverse Compton processes, and the jet, if aligned with the observer, may be detected as beamed emission and reach high apparent luminosities. 

The rate of TDEs at the centres of galaxies depends on the black hole mass and on the stellar population. In high-mass galaxies, N-body simulation \citep{2011MNRAS.418.1308B} predicted rates of 
$3.0(\pm1.4)-8.3(\pm4.2)  \times 10^{-5}\; \rm galaxy^{-1}\, yr^{-1}$ for black hole masses in the range  $10^6-10^7 M_{\odot}$.

Systematic search for bright X-ray flares in active and inactive galaxies using the {\it Rosat} All Sky Survey \citep{2002AJ....124.1308D} resulted in a rate of $1.8 \times 10^{-5}\,\rm galaxy^{-1}\, yr^{-1}$ for a peak flux detection limit $<10^{-12}\, {\rm erg}\,{\rm s}^{-1} {\rm cm}^2$ (unabsorbed). 
Excluding active galaxies, the rate dropped to $9.1 \times 10^{-6}\; \rm galaxy^{-1}\, yr^{-1}$. 
After correcting for an estimate of the effect of local obscuration
in the soft X-rays band, the corrected rate for  TDEs  increased to $1.5 \times 10^{-5}\,\rm galaxy^{-1}\, yr^{-1}$. A slightly higher rate was obtained by \cite{2014MNRAS.444.1041K}, who searched for fainter TDE candidates by correlating {\it Rosat} and {\it XMM-Newton} observations. The detection rate found in the {\it XMM-Newton} Slew Survey in non-active galaxies is  $2.3 \times 10^{-4}\; \rm galaxy^{-1}\, yr^{-1}$ \citep{2008A&A...489..543E}, for a peak flux detection limit of $2.5\times 10^{-13}\, {\rm erg} {\rm s} {\rm cm}^{-2}$. We note that signature for AGN activity was revealed by the [O  III] 5007\AA/H$\beta$ line ratios in three out of the five TDE candidates detected in the {\it XMM-Newton} slew survey \citep{2008A&A...489..543E}.

Only two TDE candidates were detected in the optical from the SDSS Stripe 82 field \citep{2011ApJ...741...73V}. \citet{2014ApJ...792...53V} injected artificial TDE flares \citep[based on the simulations by][]{2014ApJ...783...23G} in randomly selected SDSS targets and estimated an upper limit on the TDE rate of  $<2\times 10^{-4}\,\rm galaxy^{-1}\,yr^{-1}$ (90\% confidence level), in reasonable agreement with the rates determined from the X-ray observations.

The wide fields of view of the Burst Alert Telescope \citep[BAT,][]{2005SSRv..120..143B} on board  the {\it Swift} satellite and of the IBIS imager on board {\it INTEGRAL} \citep{2003A&A...411L.131U} allow   a large fraction of the sky at hard X-rays to be monitored (the complete sky every few hours for {\it Swift}/BAT). Two of the four TDEs detected at hard X-rays, Swift J164449.3+573451 \citep{2011Sci...333..199L, 2011Natur.476..421B,2011Natur.476..425Z} and Swift J2058.4+0516~\citep{2012ApJ...753...77C}, are beamed, as confirmed by radio follow-up detections and luminosities that exceed the Eddington luminosity by large factors. The two other TDEs, discovered by {\it INTEGRAL}, were interpreted as the disruption of planetary objects, either by the super-massive black hole at the center of NGC 4845 \citep{2013A&A...552A..75N} or by a white dwarf in the globular cluster NGC 6388 \citep{2014MNRAS.444...93D}.

\section{Sample selection, data extraction and filtering\label{sec:selection}}

Until now, non-relativistic candidate tidal disruption events with emission at hard X-rays were detected in nearby galaxies or in globular clusters. 
As we aim to analyse {\it Swift}/BAT lightcurves for a large sample of candidate sources, we use the most complete catalogue of galaxies within 100 Mpc together with the list of globular clusters in the Milky Way. These objects were collected in the Gravitational Waves Galaxies Catalogue \citep[GWGC,][]{2011CQGra..28h5016W}, which was built as an input for the search of gravitational waves from merging compact binaries and bursts from SNe, gamma-ray bursts, and magnetars. Included were 53\,225 galaxies with L$_{\rm B}>10^7$ L$_{\odot}$ and 150 globular clusters \citep{1996AJ....112.1487H}. All TDEs that reach luminosities $>2.84 \times 10^{44}\,{\rm erg}\,{\rm s}^{-1}$ within 100 Mpc could, therefore, potentially be detected (assuming a flare significance $>10\;\sigma$ and duration $>10$ days). This corresponds to the Eddington luminosity for a $2.3 \times 10^6$ M$_{\odot}$ black hole at 100 Mpc. 

The {\it Swift}/BAT reduction pipeline is described in \cite{2010ApJS..186..378T} and \cite{2013ApJS..207...19B}. Our pipeline is based on the BAT analysis software HEASOFT v 6.13. A first analysis was performed with the task \texttt{batsurvey} to create sky images in the eight standard energy bands using an input catalogue of 86 bright sources (that have the potential to be detected in single pointings) for image cleaning. Background images were then derived, removing all detected excesses with the task \texttt{batclean,} and the weighted average determined on a daily basis. The variability of the background was then smoothed pixel by pixel using a polynomial model with an order equal to the number of months in the data set. The BAT image analysis was then run again using these averaged background maps. The image data were then stored in a database organised by sky pixel (using a fixed sky pixel grid with pixel $< 5$ arcmin) by  accurately projecting the images on the sky pixels, whilst preserving fluxes. This database can then be used to build images, spectra, or lightcurves for any sky position.

The result of our processing was compared to the standard results presented by the {\it Swift} team (lightcurves and spectra of bright sources from the {\it Swift}/BAT 70-month survey catalogue\footnote{\texttt{http://swift.gsfc.nasa.gov/results/bs70mon/}}) and a very good agreement was found. The {\it Swift} team has corrected the spectral response derived from simulations to obtain a good fit of the very high signal-to-noise spectrum of the Crab nebula. The spectrum of the Crab nebula that we obtained shows similar deviations with respect to the simulations, but the deviations we derive are a factor of $\sim2$ smaller than those obtained by the {\it Swift} team. We corrected the response in a similar way to the {\it Swift} team, but used `fudge factors' that were a factor of $\sim2$ smaller. These modifications have no impact on the low signal-to-noise sources that we  consider in the present study.

{\it Swift}/BAT images and lightcurves were built for every object of the catalogue from early 2005 to the end of 2013. Light-curves could be extracted for any time binning and energy range. For each time bin a weighted mosaic of the selected data is first produced and the source flux is extracted, assuming a fixed source position and shape of the point spread function. The signal-to-noise of the source varies regularly because of its position in the BAT field of view with larger uncertainties observed every year. We tested various energy bands and finally extracted the lightcurves that use the complete 20-195 keV band. This energy range was used throughout the whole analysis. Removing,  for example, the highest energies did not increase the signal to noise significantly.

As the X-ray emission of tidal disruption events is expected to last for several weeks, we decided to extract lightcurves with a resolution of five  days, which is good enough to preserve the general shape of the expected flares and allows a significant reduction of the noise. 

\begin{figure}[h]
\includegraphics[width=\columnwidth]{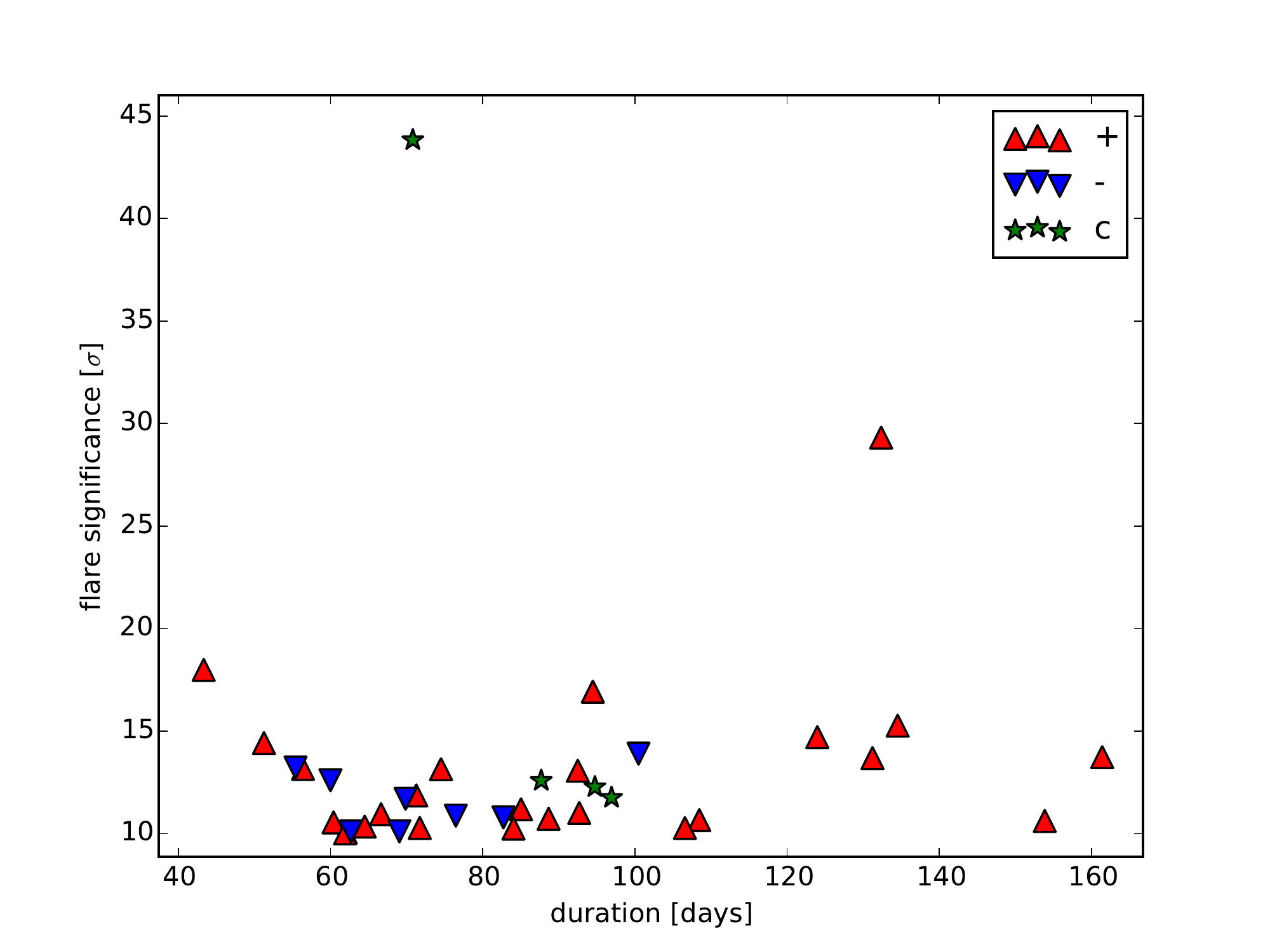}
\caption{Distribution of positive and negative flares over significance-duration space in its most significant part. Red triangles show galaxies with positive flares, green stars shows star clusters and blue triangles show negative flares in the catalogue.}
\label{fig:dist}
\end{figure}

Flare candidates were searched for using the following procedure for each source. As a first step, we searched for the most significant bin in the lightcurve, then defined a flare duration by the period of time during which the flux remained above zero and calculated the global significance of the flare by integrating the lightcurve over that period. This was repeated for all excesses found in the lightcurve and an image of the most significant flare was finally built to measure its significance accurately. One flare was therefore kept for every source for further analysis.

Because a number of the galaxies feature continuous X-ray activity, they are probably active galactic nuclei (or superimposed with a galactic source). We excluded all sources from our flare candidates  for which the time-averaged count rate, not including the most significant flare, was higher than 15\% of the standard deviation of the count rate over the complete 5-day binned lightcurve. This criteria proved to  select real persistent sources (or these contaminated by a nearby persistent source) effectively, with a total (integrated over eight years) significance $>4\;\sigma$ (actually $>10\;\sigma$ for most of them).
This meant 288 persistent sources were selected after the removal of confused flares (i.e. separated by less than 4 arcmin). Among them 115 are known as AGNs, four as galaxies, seven as galaxy clusters and nine as galactic binaries, in the BAT catalogue \citep{2013ApJS..207...19B}, where we use a maximum distance of 4 arcmin for source identification.
In the remaining 153 persistent sources, we did not find a BAT source within 4 arcmin, however, for most of them a source of the BAT catalogue is closer to the fitted position than the original GWGC galaxy, indicating a clear contamination. In only 13 cases the persistent source had a PSF shifted by less than 6 arcmin from the GWGC galaxy and no counterpart in the BAT catalogue within 100 arcmin. These cases are further discussed in section 4.1.

To understand the significance limit that should be applied to our list of flares, we used the same flare detection procedure as above on the negative lightcurves and compared the positive and negative flare distributions in a significance-duration diagram. It turned out that flares with significance $<10\;\sigma$ or duration $<40$ days have a high probability to be spurious. Additionally we excluded flares within the ellipse matching $(({\rm duration}-40)/20)^2 + (({\rm significance}-10)/5)^2 < 1$) where a few spurious negative flares could also be detected.
Those regions of the significance-duration diagram were excluded to limit the chance of finding spurious flares in the positive flare sample.

Finally we also rejected flares for which the integrated BAT images (about one degree in size) did show minima below -7 $\sigma$ (or +7 $\sigma$ for the negative flares) to eliminate high background events where  count rates were not expected statistically.

The distribution of the remaining positive and negative flares is presented in Fig.~\ref{fig:dist}. Some negative flares below 15 $\sigma$ and with a duration between 60 and 80 days are still detected, and should be taken into account in the analysis of the positive flares.

We end up with 29 positive flare candidates (25 galaxies (two galaxy clusters) and four globular clusters) and eight negative flares. 
The two galaxy clusters contained identified AGNs thus we removed those from our sample. Another two candidates correspond to the same flare, however their separation was high enough to avoid exclusion by the process described above. Thus we excluded the less significant of those two. At this point we are left with 26 candidates for further investigation.
The positive flare parameters are listed in Table~\ref{tab:par}. All fluxes and luminosities are given for the 20-195 keV energy band. The corresponding flare light curves and images are presented in Fig. \ref{fig:obj}.

\begin{table*}[t]
\caption{Flare properties and their host candidates. The flare's properties (peak epoch, duration, peak flux, significance) were measured at the centre of the PSF fit, while the distance refers to the host itself and are taken from \cite{2011CQGra..28h5016W}. In the final column we list the most likely counterpart of the flare. The order of the table is as in Sect. \ref{sec:verification}.}
\label{tab:par}
\begin{tabular}{lrrrrrl}
\hline\noalign{\smallskip}
GWGC name & $D_{L}$ & MJD & Duration & Peak flux & Significance & Counterpart \\ 
  & [Mpc] &  & [days] & [$10^{-10}\,{\rm erg}\,{\rm s}^{-1}\,{\rm cm}^{-2}$] &   &   \\ 
\noalign{\smallskip}\hline\noalign{\smallskip}
{\it Galactic objects} &  &  &  &  &  &   \\
\noalign{\smallskip}\noalign{\smallskip}
Terzan 5                 &  0.001 & 55483 & 72 & 13.27 & 45.4 & IGR J17480-2446  \\ 
Liller 1                 &  0.001 & 55812 & 86 & 2.46 & 13.3 & MXB 1730-335  \\ 
NGC 6388                 &  0.001 & 55771 & 96 & 3.23 & 12.7 &   \\ 
NGC 6293                 &  0.001 & 54583 & 93 & 3.52 & 12.1 & RX J1709.5-2639  \\ 
PGC 3079807              &  65.2 & 54798 & 132 & 25.37 & 29.9 & Sw J1539.2-6227  \\ 
PGC 016029               &  57.3 & 55240 & 95 & 5.30 & 17.0 &  KT Eri \\ 
\noalign{\smallskip}\hline\noalign{\smallskip}
{\it Gamma-ray bursts} &  &  &  &  &  &   \\
\noalign{\smallskip}\noalign{\smallskip}
NGC 3835                 &  34.1 & 53833 & 47 & 1.95 & 17.1 &  GRB 060319 \\ 
SDSS J114153.19+510158.7 &  97.5 & 55125 & 50 & 3.68 & 14.9 &  GRB 091020A \\ 
ESO 542-022              &  76.0 & 55471 & 55 & 2.91 & 13.3 &  GRB 100814A \\ 
\noalign{\smallskip}\hline\noalign{\smallskip}
{\it Relativistic TDE} &  &  &  &  &  &   \\
\noalign{\smallskip}\noalign{\smallskip}
NGC 6213                 &  78.6 & 55745 & 121 & 1.92 & 15.2 & Sw J1644+5734  \\ 
\noalign{\smallskip}\hline\noalign{\smallskip}
{\it TDE candidates} &  &  &  &  &  &   \\
\noalign{\smallskip}\noalign{\smallskip}
UGC 01791                &  67.2 & 53429 & 82 & 5.62 & 13.7 &   \\ 
PGC 015259               &  59.1 & 55238 & 60 & 3.71 & 10.4 &   \\ 
NGC 6021                 &  67.0 & 53405 & 71 & 3.31 & 10.1 &   \\ 
UGC 03317                &  20.8 & 55469 & 61 & 4.89 & 9.8 &   \\ 
PGC 170392               &  69.3 & 53730 & 69 & 4.87 & 10.6 &   \\ 
PGC 1185375              &  22.9 & 55258 & 41 & 6.69 & 19.8 &   \\ 
PGC 133344               &  97.1 & 53480 & 65 & 3.08 & 11.7 &   \\ 
PGC 1190358              &  33.0 & 55264 & 108 & 4.57 & 10.4 &   \\ 
PGC 1127938              &  76.8 & 54116 & 60 & 3.57 & 10.2 &   \\ 
\noalign{\smallskip}\hline\noalign{\smallskip}
{\it Dubious cases} &  &  &  &  &  &   \\
\noalign{\smallskip}\noalign{\smallskip}
NGC 1367                 &  17.3 & 53737 & 82 & 3.99 & 11.5 & SN 2005ke  \\ 
PGC 053636               &  26.1 & 53930 & 81 & 2.29 & 10.4 &   \\ 
SDSS J114553.33+635411.6 &  58.0 & 53828 & 70 & 1.54 & 11.1 &   \\ 
NGC 6206                 &  72.5 & 55750 & 71 & 3.22 & 13.4 &   \\ 
ESO 161-001              &  29.1 & 54073 & 90 & 2.49 & 11.3 &   \\ 
PGC 2464645              &  31.2 & 55846 & 92 & 2.70 & 12.9 &   \\ 
PGC 3241255              &  95.7 & 55214 & 65 & 2.81 & 9.9 &   \\ 
\noalign{\smallskip}\hline
\end{tabular}
\end{table*}

The point spread function of {\it Swift} BAT has a FWHM of 17 arcmin \citep{2005SSRv..120..143B}. In some cases, several galaxies from our catalogue could therefore correspond to a single flare. In these cases, we selected as a counterpart the galaxy for which the flare appeared the most significant. These cases are further discussed in Section \ref{sec:verification}.

A visual inspection did not reveal any known X-ray emitting low redshift active galaxy in the vicinity of our flare candidates. In a few cases, discussed below, optically selected AGNs \citep{2010A&A...518A..10V} could potentially contaminate if they were to emit at hard X-rays.

\section{Flare candidate analysis \label{sec:verification}}

We analysed  each of the 26 flare candidates manually. As eight candidates were found in the `negative' space, we expect some of the `positive' candidates to be of instrumental origin.

\begin{table*}[th]
\caption{Host name and position (Columns 1-5) and flare position (Columns 6-7) and uncertainty (Column 8) from the fit of the point spread function. The last column lists the differences between the GWGC candidate position and the PSF fit centre.}
\label{tab:psf}
\begin{tabular}{lrrrrrrrr}
\hline\noalign{\smallskip}
GWGC Name & Galactic $l$ & Galactic $b$ & RA & Dec & PSF RA & PSF Dec & Error & Offset \\ 
  & [deg] & [deg] & [deg] & [deg] & [deg] & [deg] & [arcmin] & [arcmin] \\ 
\noalign{\smallskip}\hline\noalign{\smallskip}
{\it Galactic objects} &  &  &  &  &  &   \\
\noalign{\smallskip}\noalign{\smallskip}
Terzan 5                 &    3.839 & 1.687 &  267.020 & -24.779 & 267.038 & -24.782 & 0.7 & 1.0 \\ 
Liller 1                 &  354.841 & -0.160 &  263.352 & -33.389 & 263.341 & -33.423 & 1.9 & 2.1 \\ 
NGC 6388                 &  345.556 & -6.737 &  264.070 & -44.735 & 264.095 & -44.691 & 2.0 & 2.8 \\ 
NGC 6293                 &  357.620 & 7.834 &  257.543 & -26.582 & 257.398 & -26.650 & 1.1 & 8.8 \\ 
PGC 3079807              &  321.085 & -5.726 &  235.022 & -62.495 & 234.797 & -62.470 & 0.8 & 6.4 \\ 
PGC 016029               &  208.032 & -32.068 &   71.949 & -10.235 &  71.990 & -10.162 & 1.3 & 5.0 \\ 
\noalign{\smallskip}\hline\noalign{\smallskip}
{\it Gamma-ray bursts} &  &  &  &  &  &   \\
\noalign{\smallskip}\noalign{\smallskip}
NGC 3835                 &  137.527 & 55.059 &  176.020 & 60.120 & 175.946 & 60.154 & 1.3 & 3.0 \\ 
SDSS J114153.19+510158.7 &  147.124 & 62.703 &  175.472 & 51.033 & 175.629 & 51.071 & 1.5 & 6.4 \\ 
ESO 542-022              &  169.255 & -76.835 &   22.816 & -17.699 &  22.615 & -17.880 & 0.9 & 15.8 \\ 
\noalign{\smallskip}\hline\noalign{\smallskip}
{\it Relativistic TDE} &  &  &  &  &  &   \\
\noalign{\smallskip}\noalign{\smallskip}
NGC 6213                 &   87.118 & 39.815 &  250.405 & 57.815 & 251.137 & 57.732 & 0.6 & 23.9 \\ 
\noalign{\smallskip}\hline\noalign{\smallskip}
{\it TDE candidates} &  &  &  &  &  &   \\
\noalign{\smallskip}\noalign{\smallskip}
UGC 01791                &  145.616 & -30.698 &   34.973 & 28.248 &  34.977 & 28.197 & 1.7 & 3.0 \\ 
PGC 015259               &  199.738 & -33.581 &   67.341 & -4.760 &  67.344 & -4.816 & 2.1 & 3.4 \\ 
NGC 6021                 &   28.259 & 45.576 &  239.378 & 15.956 & 239.385 & 15.957 & 2.5 & 0.4 \\ 
UGC 03317                &  139.350 & 20.771 &   83.407 & 73.724 &  83.234 & 73.688 & 2.2 & 3.6 \\ 
PGC 170392               &   45.287 & -54.305 &  336.693 & -15.023 & 336.722 & -15.048 & 2.7 & 2.2 \\ 
PGC 1185375              &  359.212 & 48.971 &  225.960 & 1.127 & 225.933 & 1.091 & 1.2 & 2.7 \\ 
PGC 133344               &   17.207 & -48.817 &  325.733 & -30.133 & 325.737 & -30.081 & 2.4 & 3.1 \\ 
PGC 1190358              &  359.815 & 48.778 &  226.369 & 1.292 & 226.391 & 1.303 & 2.5 & 1.5 \\ 
PGC 1127938              &  138.276 & -63.103 &   19.736 & -1.053 &  19.681 & -1.137 & 2.1 & 6.0 \\ 
\noalign{\smallskip}\hline\noalign{\smallskip}
{\it Dubious cases} &  &  &  &  &  &   \\
\noalign{\smallskip}\noalign{\smallskip}
NGC 1367                 &  218.954 & -53.347 &   53.755 & -24.933 &  53.997 & -25.047 & 1.6 & 14.8 \\ 
PGC 053636               &  358.020 & 49.199 &  225.263 & 0.708 & 225.416 & 0.669 & 1.8 & 9.5 \\ 
SDSS J114553.33+635411.6 &  134.491 & 51.729 &  176.472 & 63.903 & 176.702 & 63.832 & 1.7 & 7.4 \\ 
NGC 6206                 &   88.194 & 39.831 &  250.033 & 58.617 & 250.238 & 58.585 & 1.4 & 6.7 \\ 
ESO 161-001              &  266.355 & -26.674 &   95.126 & -57.495 &  94.956 & -57.548 & 2.2 & 6.3 \\ 
PGC 2464645              &  101.430 & 59.664 &  211.342 & 54.270 & 211.232 & 54.159 & 1.7 & 7.7 \\ 
PGC 3241255              &  217.829 & -65.764 &   40.067 & -26.550 &  40.198 & -26.507 & 2.2 & 7.5 \\ 
\noalign{\smallskip}\hline
\end{tabular}
\end{table*}

This verification started by checking the image of the field that was obtained during each flare. The point spread function was fitted to the image using a 2D Gaussian profile with a width $\sigma=8.2$ arcmin (with \texttt{mosaic\_spec} from the OSA toolkit\footnote{http://isdc.unige.ch/integral/analysis}). The flare central coordinates are listed in columns \texttt{PSF RA} and \texttt{PSF Dec} of Table~\ref{tab:psf}, together with the positional uncertainty (\texttt{Error} column) at 90\% confidence. The angular distance between the most likely counterpart derived in Sect. \ref{sec:selection} and the centre of the PSF is given in the column \texttt{Offset}. The error circles that result from the PSF fitting are indicated as the inner white dotted circles in Fig.~\ref{fig:obj}, while the outer white-black dashed circles illustrate the radius of the PSF. When the most likely counterpart size is available it is indicated with a magenta circle in Fig. \ref{fig:obj}. The difference between the counterparts and best fit positions vary between 0.4 and 24 arcminutes. Differences larger than 10 arcminutes are most likely indicating a spurious association. Very large ($> 20$ arcmin) differences indicate that the flare image is far from a point source and most likely related to instrumental or image cleaning problems.

Also indicated in the flare images are the positions of other sources from the Gravitational Waves Galaxies Catalogue (red open circles) and from several other catalogues (the 13th edition of the Catalogue of Quasars and Active Galactic Nuclei \citep[cyan stars;][]{2010A&A...518A..10V}, the 7-year {\it Swift}-XRT point source catalogue \citep[yellow squares;][]{2013A&A...551A.142D}, the {\it Swift}-BAT survey of Galactic sources \citep[green diamonds;][]{2010ApJ...721.1843V}, the {\it Fermi} LAT second source catalogue \citep[blue triangles;][]{2012ApJS..199...31N}, the second {\it Fermi}/GBM GRB catalogue \citep[blue crosses;][]{2014ApJS..211...13V}, the list of supernovae observed by {\it Swift}/XRT \citep[cyan x;][]{2012ApJ...748L..29R}, the {\it Swift} X-ray observations of classical novae II \citep[magenta x;][]{2011ApJS..197...31S}, the {\it Swift}/BAT hard X-ray transient monitor catalogue \citep[magenta stars;][]{2013ApJS..209...14K}, the second {\it Swift} BAT GRB catalogue \citep[red +;][]{2011ApJS..195....2S}, the 70 month {\it Swift}-BAT all-sky hard X-ray survey \citep[red x;][]{2013ApJS..207...19B} obtained through the Vizier service\footnote{http://vizier.u-strasbg.fr/viz-bin/VizieR}.

As very bright sources might affect the whole field of view of the BAT instrument and the source cleaning might be imperfect, we also checked whether our flare candidates occurred when any of the BAT bright sources were present within 60 degrees and flaring at the same time. This check was performed by comparing the flare candidate lightcurve with those of the very bright sources. Specific discussions are only required  in a few cases and mentioned below.

When a low significance flare occurs in regions with significant background, the fitting of the instrument point-spread function to the flare excess can be unreliable and provide inaccurate results, in particular on the uncertainty of the flare position. These cases are discussed specifically. 

\begin{figure}
\includegraphics[width=0.49\textwidth]{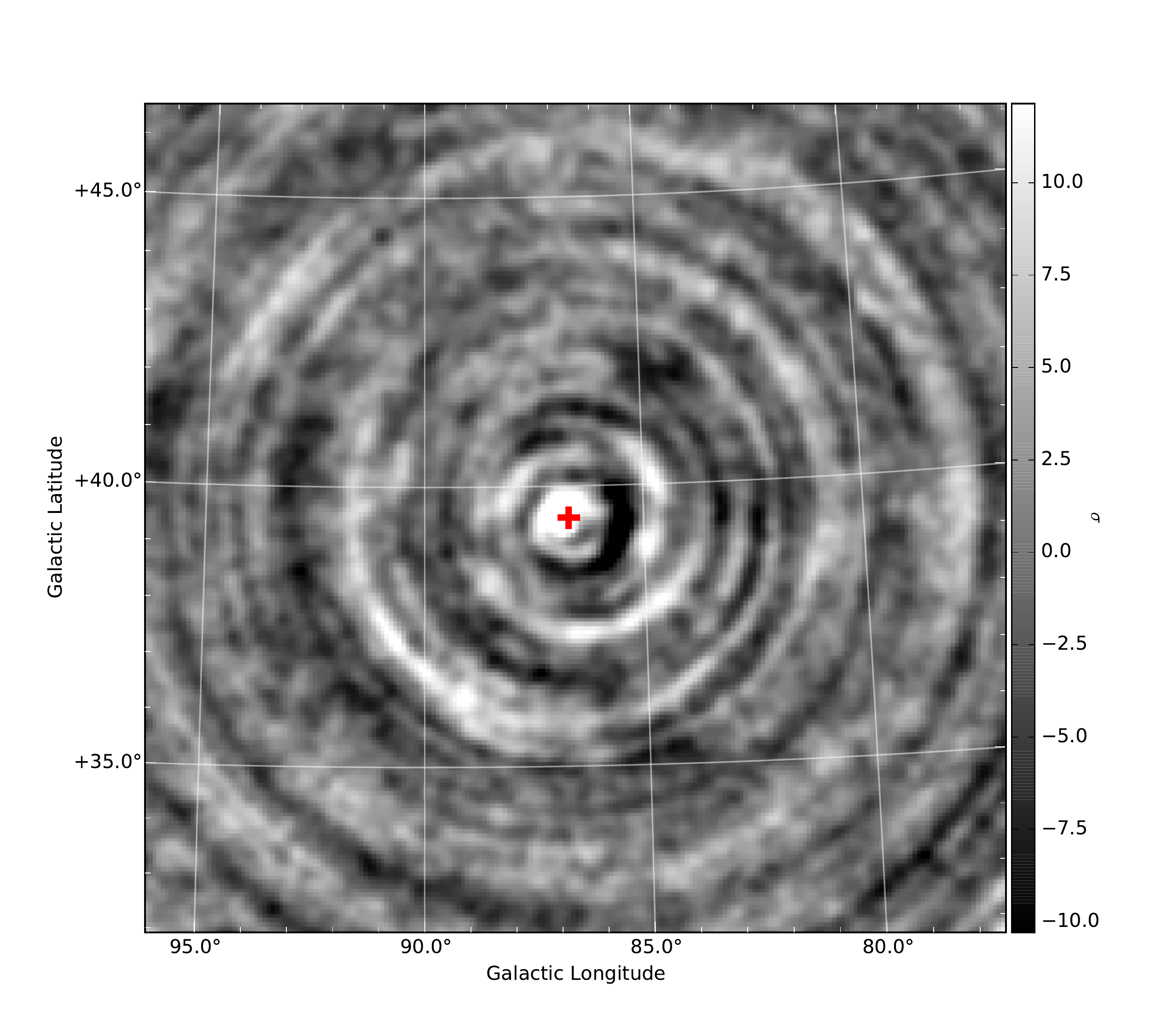}
    \caption{Illustration of the coding noise rings built during star observations (here for an exposure of 121 days). This image is centred on GRB 110328A, which is indicated by the red cross.}
\label{fig:rings}
\end{figure}

{\it Swift} observed gamma-ray burst afterglows for a long period of time with BAT pointing in a fixed direction and its field of view rotating on the sky because of Solar constraints. During this rotation, coding noise accumulated along arcs in the mosaic image at a level of typically $(6-7)\;\sigma$ for an exposure of five days (see an example in Fig. \ref{fig:rings}). Such arcs are easy to spot, by analysing a wide field of view (at least three degrees), and some of our flare candidates do correspond to local maxima of an arc located by chance at the position of a GWGC object. These cases are discussed in the following subsections.

In the following subsections, we describe each individual flare, sorted into four categories that depend on their most likely physical explanation.
Only the sources listed in Sect. 3.4 are considered as real TDE candidates.
Fig. \ref{fig:obj} displays the field images and the BAT lightcurves, extracted at the positions of the flare, as derived from the point-spread function best fit.


\subsection{Galactic objects}

In this subsection we describe seven flare candidates, which  likely originate in the Milky Way and were emitted by sources located either in globular clusters or in the Galactic plane.
None of these sources are considered as a TDE candidate.

\subsubsection{Terzan 5 and IGR J17480-2446} 

This flare is very significant (45.4 $\sigma$) and its image point-like. The flare onset occurred after  2010 Oct 01, peaked on 2010 Oct 14, and probably lasted for about 72 days (up to 108 days). 
The flare, with a position consistent with the centre of the globular cluster Terzan 5, was detected by {\it INTEGRAL} \citep{2010ATel.2919....1B} and allowed the discovery of the 
new accreting pulsar IGR J17480-2446 \citep{2011A&A...526L...3P}. The second brightest flare in Terzan 5 peaked on 2012 July 18 and corresponds to a Type-I burst, which led to 
the discovery of the new accreting neutron star Terzan 5 X-3 \citep{2014ApJ...780..127B}.


\subsubsection{Liller 1} 

A significant flare was detected at the position of the globular cluster Liller 1 from 2011 July 31 to 2011 Oct 25, peaking on 2011 Sept 07, with a total duration of approximately 86 days. 
It was also detected by {\it INTEGRAL} and attributed to the rapid burster MXB 1730-335 \citep{2011ATel.3646....1C}. 
The BAT flare image is centred 2.1 arcmin from the rapid burster. Flares from this source are regularly detected  with {\it Swift}/BAT. 
The low mass X-ray binary 4U 1728-34, located 30 arcmin away, is much brighter but did not contaminate the flare significantly. 

\subsubsection{NGC 6388} 

A significant flare was detected at the position of the globular cluster NGC 6388 from 2011 July 16 to 2011 Oct 20, peaking on 2011 Sept 28, with a total duration of approximately 96 days. It was also detected by {\it INTEGRAL} and attributed to the disruption of an icy planet by a white dwarf~\citep{2014MNRAS.444...93D}. The low mass X-ray binary 4U 1735-44, located 31 arcmin away, is much brighter but did not contaminate the flare significantly. 

\subsubsection{NGC 6293 and RX J1709.5-2639} 

A significant flare was detected at the position of the globular cluster NGC 6293 from 2008 March 08 to 2008 June 09, peaking on 2008 April  26, with a total duration of approximately 93 days.
The BAT flare image is centred 8.8 arcmin away from the centre of NGC 6293, which is significant, but inside the globular cluster. The source shows recurrent flares in the BAT lightcurve, which are also detected by ISGRI and JEM-X on board {\it INTEGRAL}. The most accurate source position, derived by  accumulating all JEM-X data, is RA $=257.378$; DEC $=-26.656$; $\Delta=0.7$ arcmin, which corresponds to that of the low-mass X-ray binary RX J1709.5-2639 \citep{2013ApJ...767L..31D}.

\subsubsection{PGC 3079807 and Swift J1539.2-6227} 

A significant flare was detected at the position of the galaxy PGC 3079807 from 2008 Nov 20 to 2009 April 01, peaking on 2008 Nov 27, with a total duration of approximately 132 days.
The BAT flare image is centred 6.4 arcmin away from PGC 3079807, which is significant and matches (within 0.02 arcmin) the position of LMXB Swift J1539.2-6227, marked with a magenta star in Fig.~\ref{fig:obj}. Our flare also matches in time with the discovery of Swift J1539.2-6227 \citep{2011ApJ...735..104K}. It is not linked with PGC 3079807.



\subsubsection{PGC 016029 and KT Eri} 

A significant flare was detected at the position of the galaxy PGC 016029 from 2010 Jan 01 to 2010 April 06, peaking on 2010 Feb 12, with a total duration of approximately 95 days.
The BAT flare image is centred 5.0 arcmin away from PGC 016029, which is significant and matches (within 1.3 arcmin) the position of the classical novae KT Eri \citep{2011ApJS..197...31S},
marked in Fig.~\ref{fig:obj} with a magenta x. KT Eri erupted in November 2009 and became a super soft source in December \citep{2010ATel.2392....1B} and reached an X-ray maximum in January 2010 \citep{2011ApJS..197...31S}. The hard X-ray flare detected by BAT is, therefore, not related to PGC 016029.



\subsection{Gamma-ray burst afterglows}

In this subsection, we list three flare candidates that correspond to known gamma-ray bursts. These gamma-ray bursts (GRB) are detected by chance and are not related to our sample of nearby galaxies. The BAT images of GRB are affected by circular artefacts that result from the stacking of coding noise, which is  generated by many follow-up XRT observations that point to the GRB position, but rotated with respect to each other. In the cases reported here, XRT pointings took place for a significant amount of the flare duration and
the BAT flare duration and fluxes are completely spurious. None of these sources are considered  a TDE candidate.

\subsubsection{NGC 3835 and GRB 060319} 

This flare, which occurred between 2006 March 16 and 2006 April 30, featured a constant flux during the first 20 days and then faded away. Its total duration is between 47 and 135 days. The flare is shifted from NGC 3835 by 3.0 arcmin, which is outside of the galaxy. 

The gamma-ray burst GRB 060319 \citep{2011ApJS..195....2S} occurred on 2006 March 19 for 8.9 s (shown as a vertical dashed line on the lightcurve in Fig. \ref{fig:obj}), when we detected a big rise in the BAT count rate. The GRB position is marked on the image in Fig.~\ref{fig:obj} with a red cross which is within 15 arcmin of our flare position.



\subsubsection{SDSS J114153.19+510158.7 and GRB 091020} 

This flare, which occurred between 2009 Sept 28 and 2009 Nov 18, featured a maximum on 2009 Oct 10-21 $(\pm 2.5 {\rm d})$ and decayed afterwards. 
Its total duration is between 20 and 50 days. The flare is shifted from SDSSJ114153.19+510158.7 by 6.4 arcmin. 

The gamma-ray burst GRB 091020A \citep{2011ApJS..195....2S, 2014ApJS..211...13V} occurred on 2009 Oct 20 (shown as a vertical dashed line on the lightcurve in Fig. \ref{fig:obj}), when we detected a rise of the BAT count rate. The GRB position is marked on the image in Fig.~\ref{fig:obj} with a red cross, which is within 6.7 arcmin from our flare position. The difference between the flare and the GRB positions comes from the different periods of time used for the image integration.



\subsubsection{ESO 542-022 and GRB 100814A} 

This flare, which occurred between 2010 Aug 10 and 2010 Oct 04, has a rather undefined shape because of a limited signal to noise of $13\;\sigma$ and a lack of features. Its total duration is $\sim 55$ days. The flare is shifted from ESO 542-022 by 15.8 arcmin. 

The gamma-ray burst GRB 100814A \citep{2011A&A...531A..20G} occurred on 2010 Aug 14 (shown as a vertical dashed line on the lightcurve in Fig. \ref{fig:obj}), which is consistent with our flare lightcurve. The GRB position is marked on the image in Fig.~\ref{fig:obj} with a red cross, which is within 9.7 arcmin from our flare position. The difference between the flare and GRB positions comes from the different periods of time used for the image integration.




\subsection{Relativistic TDE candidates}

One of the relativistic TDE candidate detected by {\it Swift} occurred close to a GWGC galaxy and was therefore detected by our analysis.

\subsubsection{NGC 6213 and GRB 110328A/Swift J164449.3+573451\label{ssec:reltde}} 

This flare, detected between 2011 March 26 and 2011 July 25, is very noisy and the image reveals that the emission is centred 24 arcmin
away from NGC 6213, on an unidentified persistent source (recognised as such by the criteria described in Sect. \ref{sec:selection}.)
The flare is not associated with NGC 6213.

The gamma-ray burst GRB 110328A, which occurred on 2011 March 28 (shown as a vertical dashed line on the lightcurve in Fig. \ref{fig:obj}),  
is consistent with the onset of the flare lightcurve. The GRB position determined by {\it Swift}/XRT and indicated with a red cross in Fig.~\ref{fig:obj} is within $9.3 \pm0.6$ arcmin from the flare's best-fit position. The flare image shows extended background emission, related to the circular artefacts,  and, therefore, the flare's positional uncertainty is underestimated and the lightcurve is uncertain.

GRB 110328A/Swift J164449.3+573451 is not an ordinary gamma-ray burst. It was dubbed the longest GRB~\citep{2012MNRAS.419L...1Q} and was very bright ($> 10^{47}\, {\rm erg}\,{\rm s}^{-1}$). XRT follow-up revealed a new X-ray source but no UVOT detection was reported. This burst occurred in a small star-forming galaxy located at redshift $z=0.35$~\citep{2011Sci...333..199L}. Its central black hole mass was constrained from X-ray and radio observations as log $M = 5.5 \pm 1.1 M_{\odot}$~\cite{2011ApJ...738L..13M}. Several explanations were proposed to discuss the nature of this super long burst and, among them, a tidal disruption/fall-back accretion onto a massive black hole with collimated relativistic jet formation \citep{2011Natur.476..421B, 2011Sci...333..203B, 2011Natur.476..425Z,2011ApJ...734L..33S}. \cite{2011ApJ...743..134K} addressed short-term variability during the X-ray flare by postulating a white dwarf disruption scenario, however the time-scale and fall-back rate analysis \citep{2011ApJ...742...32C} suggested a tidal obliteration event. Alternative mechanisms, such as the collapse of the galactic nucleus \citep{2012GrCo...18..232D} or the emission from a millisecond pulsar \citep{2011ApJ...734L..33S}, were also considered.

\subsection{TDE candidates}

This section presents flares for which we did not find any clear spurious counterpart or contaminant, and which are positionally consistent with one of the GWGC galaxy. This sample brings together  our best TDE candidates.
TDE light curves are parametrized by a variability factor that is defined as the ratio between the luminosity of the flare's peak $L_{\rm peak}$  and the average luminosity over the quiescence period $\bar{L}_{\rm quiescence}$.

\subsubsection{UGC 01791} 

This flare occurred between 2005 Jan 09 and 2005 April 01, with the most significant peak at 2005 Feb 27. Its duration could be between 50 and 130 days. The flare is shifted from UGC 01791 by $3.0\pm1.7$ arcmin and therefore, could come from that galaxy. We did not find any other source within the flare's PSF radius. The closest XRT source is well outside of the PSF in the flare image. There is no circular artefact on a large-scale image.

Assuming that UGC 01791 is the real counterpart, the peak luminosity and the energy emitted during the flare would be $3 \times 10^{44}\, {\rm erg}\,{\rm s}^{-1}$ and $7 \times 10^{50} $ ergs, respectively. The variability factor is $>47$.

\subsubsection{PGC 015259} 

This flare, detected between 2010 Feb 08 and 2010 April 09, features an initial increase with a slow decline over a period of 60 days.
The flare is shifted from PGC 015259 by $3.4\pm2.1$ arcmin and could come from that galaxy. We did not find any other source within the flare PSF radius and there is no circular artefact on a large-scale image.

Assuming that PGC 015259 is the real counterpart, the peak luminosity and the energy emitted during the flare would be $1.55 \times 10^{44}\, {\rm erg}\,{\rm s}^{-1}$ and $3 \times 10^{50}$ ergs, respectively. The variability factor is $>30$.


\subsubsection{NGC 6021} 

This flare was detected between 2005 Jan 02 and 2005 March 17 and is featureless. Its duration could be 71 days but this is very approximate.
The flare is shifted from NGC 6021 by $0.4\pm2.5$ arcmin, and   is, therefore, well centred on that galaxy. 

There is an other GWGC galaxy within the PSF radius, NGC 6018, but we rejected it because it lies outside of the flare centre error circle and because the flare would be less significant for that position than our detection limit. There is no circular artefact on a large-scale image.

Assuming that NGC 6021 is the real counterpart, the peak luminosity and the energy emitted during the flare would be $1.78 \times 10^{44}\, {\rm erg}\,{\rm s}^{-1}$ and $5 \times 10^{50} $ ergs, respectively. The variability factor is $>28$.


\subsubsection{UGC 03317} 

This flare, detected between 2010 Sept 22 and 2010 Nov 22, features a maximum on 2010 Sept 29 and a decline that may last up to 61 days.
The flare is shifted from the centre of UGC 03317 by $3.6\pm2.3$ arcmin. We did not find any other source within the flare's PSF radius. There is no circular artefact on a large scale image.

Assuming that UGC 03317 is the real counterpart, the peak luminosity and the energy emitted during the flare would be $2.53 \times 10^{44}\, {\rm erg}\,{\rm s}^{-1}$ and $4 \times 10^{49}$ ergs, respectively. The variability factor is $>40$.


\subsubsection{PGC 170392} 

This flare, detected between 2005 Nov 28 and 2006 Jan 19, is quite noisy as the source reached the border of the BAT field of view during or after the flare. The flare position is shifted from PGC 170392 by $2.3\pm2.7$ arcmin. We did not find any other source within the flare's PSF radius and there is no circular artefact on a large-scale image.

Assuming that PGC 170392 is the real counterpart, the peak luminosity and the energy emitted during the flare would be $2.8 \times 10^{44}\, {\rm erg}\,{\rm s}^{-1}$ and $5 \times 10^{50} $ ergs, respectively. The variability factor is $>35$.

\subsubsection{PGC 1185375} 

This flare, detected between 2010 Feb  03 and 2010 Feb 16, is highly significant and had a duration of 41 days.
The significance maximum in the flare image is shifted from PGC 1185375's centre position by $2.7\pm1.2$ arcmin.
However,  because the elongated background in this region significantly affects the result of the point-spread function fit, the
GWGC galaxy and the flare could still be related. 
There is no clear contaminating source around, no point source detected by XRT, and no circular artefact on a large-scale image.

There is an other GWGC galaxy at the border of the PSF radius, NGC 5831, but we rejected it as it lies even further outside of the flare centre error circle and because the flare would be less significant for that position.
AGNs lying within the PSF radius are too distant ($z>0.85$) to  contribute to this emission effectively.  The elongated structure is probably instrumental.

If we assume that the flare was emitted by PGC 1185375, the peak luminosity and the energy emitted during the flare would be $4.2 \times 10^{43}\, {\rm erg}\,{\rm s}^{-1}$ and $10^{50}$ ergs. The variability factor is $>51.5$.


\subsubsection{PGC 133344} 

This flare, detected between 2005 March 17 and 2005 May 21, lasted for about 65 days.
The significance maximum in the flare image is shifted from PGC 133344's centre position by $3.1\pm2.4$ arcmin. 
There is a possible second close-by weak BAT source in the field (at more negative latitude and similar longitude, about 20 arcmin from the flare position) that could easily explain this shift in position.

If we assume that the flare was emitted by PGC 133344, the peak luminosity and the energy emitted during the flare would be $3.5 \times 10^{44}\, {\rm erg}\,{\rm s}^{-1}$ and $1 \times 10^{51}$ ergs. The variability factor is $>14$.

The second GWGC galaxy within the PSF radius is less significant (9 vs 11 sigma), more distant and further away from the PSF centre, and shows a shorter flare. AGNs within the PSF circle are at high redshifts (z>1.6).
We note that the optically identified quasars 2QZ J214250-2946, at redshift 2.8, lies 1.2 arcmin away from our flare candidate and PGC 133344 \citep{2010A&A...518A..10V}. 


\subsubsection{PGC 1190358} 

This flare, detected between 2009 Dec 29 and 2010 April 16, lasted for about 108 days with a clear maximum on 2010 March 08.

The significance in the flare image matches PGC 1190358 with a distance of $1.5\pm2.5$ arcmin. The extended image does not show circular artefacts.

If we assume that the flare was emitted by PGC 1190358, the peak luminosity and the energy emitted during the flare would be $6 \times 10^{43}\, {\rm erg}\,{\rm s}^{-1}$ and $2 \times 10^{50}$ ergs. The variability factor is $>34$.

\subsubsection{PGC 1127938} 

This flare, detected between 2007 Jan 08 and 2007 March 09, lasted for about 60 days with a peak on 2007 Jan 15.

The significance maximum in the flare image is shifted from PGC 1127938's centre position by $6.0\pm2.1$ arcmin. There is no contaminant nor circular artefact in the extended image.

If we assume that the flare was emitted by PGC 1127938, the peak luminosity and the energy emitted during the flare would be $2.5 \times 10^{44}\, {\rm erg}\,{\rm s}^{-1}$ and $4\times 10^{50}$ ergs. The variability factor is $>29$.

\subsection{Dubious cases}

Here we list the remaining flares. They have a low signal to noise, and the flare images do not show a point-like excess.
None of these sources are considered to be TDE candidates.

\subsubsection{NGC 1367} 

A significant flare was detected close to the galaxy NGC 1367 from 2005 Nov 08 to 2006 Jan 29, peaking on 2006 Jan 01, with a total duration of approximately 82 days. The BAT flare image is centred 14.8 arcmin away from NGC 1367, so does not match with the galaxy. 

The gamma-ray burst GRB 090909B  \citep{2014ApJS..211...13V} occurred in the same region as our BAT flare but occurred four years later and so does not match.

The supernovae Ia SN 2005ke which exploded on 2005 Nov 06 $\pm$ 2 days (with optical and X-ray maxima occurring $\sim 20$ and respectively $\sim 45$ days later \citep[][]{2006ApJ...648L.119I}) is 14 arcmin away from the BAT flare but these two events coincide in time. The large scale image is affected by a circular artefact centred on SN 2005ke, we must conclude that this flare is an artefact from the observations of SN 2005ke.

\subsubsection{PGC 053636} 

This flare, detected between 2006 July 11 and 2006 Sept 30, features an early maximum and a slow decrease over 81 days.
The lightcurves shows very large error bars after the flare when the source moved close to the border of the BAT field of view
and a gap near the beginning because two months of data were missing from our analysis for technical reasons.

The flare is shifted from PGC 053636 by $9.5\pm 1.8$ arcmin so they are not related to each other. There are several optically 
selected AGNs close by but no known X-ray source in the XRT catalogue. The flare around PGC 053636 originates from a large ring-like structure (like the one shown in Fig. \ref{fig:rings}), which is centred on GRB 060814, and which is completely unrelated and separated by more than 20 degrees from PGC 053636.


\subsubsection{SDSS J114553.33+635411.6} 

This flare, detected between 2006 Feb19 and 2006 April 30, features a rather flat lightcurve during about 70 days and a rather low significance. The flare could be longer as it is immediately followed by a period of two months during which data were missing in our analysis for technical reasons. The flare is shifted from SDSS J114553.33+635411.6 by $7.4\pm1.7$ arcmin. The extended image shows circular structures, the flare is an artefact resulting from the follow-up observations of GRB 060319 located 3.9 degrees away.



\subsubsection{NGC 6206} 

This flare, detected between 2011 May 11 and 2011 July 20, lasted for about 71 days and featured a peak on 2011 July 07.

The significance maximum in the image is shifted from NGC 6206's centre position by $6.7\pm1.4$ arcmin. Circular artefacts indicate that this flare resulted from XRT follow-up observations of GRB 110328A/Swift J164449.3+573451, located 1.2 degrees away.


\subsubsection{ESO 161-001} 

This flare, detected between 2006 Sept 17 and 2006 Dec 16, lasted for about 90 days.

The flares does not look at all like a point source and is an artefact, which results from XRT pointings on GRB 060729, located 4.8 degrees away.


\subsubsection{PGC 2464645} 

This flare, detected between 2011 Aug 19 and 2011 Nov 19, lasted for about 92 days and featured a peak on 2011 Oct 12.
It was found to match with  PGC 2464645 and SDSS J140357.26+542228.6 (these were not associated with 
  the deduplication algorithm, which uses a smaller distance as criteria).

The significance maximum in the image is shifted from PGC 2464645's centre position by $7.7\pm1.7$ arcmin.
Near to PGC 2464645, many XRT pointing observations was conducted within M101 galaxy probably as a result of SN 2011fe follow-up. SN 2011fe was discovered on 24 Aug 2011 in M101 at a position shifted by 17.5 arcminutes from the detected flare position.


\subsubsection{PGC 3241255} 

This flare, detected between 2009 Nov 21 and 2010 Jan 25, lasted for about 65 days.
The significance image is elongated and does not correspond to a point source. This flare is an artefact  that results from follow-up observations of GRB 091127, located 8.3 degrees away.

\section{Discussion\label{Sect:Disc}}

In this work we focused on the search for flares produced by TDEs that originate in non-active galaxies. It was,  thus, important to identify and remove persistent sources. 
As a result we found five new persistent sources, which were not known as active galaxies previously and discuss them in Sect. \ref{sec:persist}. 
The four flares that originate from globular clusters are discussed in Sect. \ref{sec:gc}. 

Most flares were detected close to non-active GWGC galaxies. Among them, three correspond to gamma-ray burst afterglows that are located by chance close to NGC 3835, SDSS J114153.19+510158.7, and ESO 542-022. The nine remaining TDE candidates are discussed in Sect. \ref{sec:tde}.

\subsection{New persistent hard X-ray sources \label{sec:persist}}

The removal of persistent sources is described in Sec.~\ref{sec:selection}, where persistence is defined as when the averaged count rate (excluding the most significant flare) divided by its standard deviation is larger than 0.15. In this section we take a closer look at the 153 persistent sources not previously known as active galactic nuclei. For the majority of them we found that a source of the BAT catalogue was closer to the X-ray source PSF centre than the GWGC galaxy. In only 13 cases the persistent source had a PSF shifted by less than six arcmin and no counterpart in the BAT catalogue within 100 arcmin. 
The reason that we found new sources not listed in \cite{2013ApJS..207...19B} could be explained by the longer time period for the data in our analysis, meaning higher significance for persistent sources.
We checked other X-ray catalogues but did not find any obvious contaminant.
In addition, we checked for  contaminants in the 13th edition of the  Quasars and Active Galactic Nuclei Catalogue \citep{2010A&A...518A..10V}. In five cases, our targets are close enough to known optical AGNs (UGC  3255, Zw 022.021, ESO 500-G34, ESO 439-G09, and UM 163), which are not listed in the BAT catalogue. 

We considered the remaining eight persistent sources as AGN candidates and rejected four of them as being dubious or contaminated for the following reasons. Several circular artefacts, as presented in Fig.~\ref{fig:rings}, affect the  lightcurve of UGC 512 and mean that this AGN  is probably  a spurious candidate. As PGC 2793378 was considered  a candidate for an obscured galaxy cluster \citep{1999A&A...352...39W} and as a {\it Fermi} AGN, 2FGL J1629.6-6141, lies 4 arcmin away, this AGN candidate is also dubious. Finally the galaxies SDSS J211238.66+112822.8 and PGC 022813 should also be disregarded as they may be contaminated by the AGN UGC 11700 and by the galaxy cluster RX J0808.7+6731, respectively.

We investigated the light curves of the four remaining  AGN candidates using different time bins and coding fractions. Two sources (ESO 385-014 and PGC 070775) could be excluded as their light curves were affected by coding noise and showed enhancements only at specific time periods. Two persistent objects remained: ESO 315-020 and ESO 153-003. The significance of the corresponding excesses in the total BAT image is 13.2 and 4.8, respectively. Thus the last one is more uncertain, however an XRT point source was also detected near the position of ESO 315-020. We conclude that these two GWGC galaxies are likely to be new active galactic nuclei.

The study of ESO 315-020 is difficult because it is contaminated by a bright foreground star \citep{2000MNRAS.313..800D} and  because  it is considered as an edge-on spiral galaxy \citep{2002MNRAS.334..646K,2005MNRAS.358..503K}, which makes the detection of activity more difficult. Optical observations of ESO 153-003 revealed a dust lane along its major axis which, together with a radial velocity measurement along the minor axis, allowed for its classification as an edge-on S0 galaxy \citep{2006MNRAS.371..633H}.

\subsection{Globular cluster\label{sec:gc}}

We also detected four flares at the positions of globular clusters included in the GWGC catalogue. The most significant flare was detected in Terzan 5 and most probably originated from the LMXB/NS EXO 1745-248. The second flare, detected in Liller 1, could be related to the rapid burster MXB 1730-335 \citep{2001AJ....122.2627H}. The third source is the TDE candidate that was discovered by~\cite{2014MNRAS.444...93D} and modelled as the disruption of an icy planet by a white dwarf. Our last candidate originated within the angular size of the cluster NGC 6293 and is most probably associated with the low mass X-ray binary RX J1709.5-2639. If this association is correct, the flare luminosity would be $4.2 \times 10^{36}\, {\rm erg}\,{\rm s}^{-1}$.

\subsection{TDE candidates\label{sec:tde}}

Our sample contained 26 flare candidates close to GWGC galaxies. After excluding all cases showing quite clear indications of contamination,  such as presence of GRB afterglows, supernova, classical nova, Galactic binaries, and dubious sources, we are left with nine candidates. These flares occurred near the galaxies UGC 01791, PGC 015259, NGC 6021, UGC 03317, PGC 170392, PGC 1185375, PGC 133344, PGC 1190358, and PGC 1127938.


If we assume that all these nine flares originated from tidal disruption events, the TDE rate, above the BAT detection limit, could be estimated as

\begin{equation}
\dot{N}_{\rm TDF} = \frac{9\; \rm events}{52874\; {\rm galaxies} \times 8\; {\rm years}} \approx 2.1 \times 10^{-5}\; \rm yr^{-1} galaxy^{-1}.
\end{equation}

This should formally be considered as an upper limit, as some of these flares could have a different origin.

\begin{figure}
\centering
\includegraphics[width=0.49\textwidth]{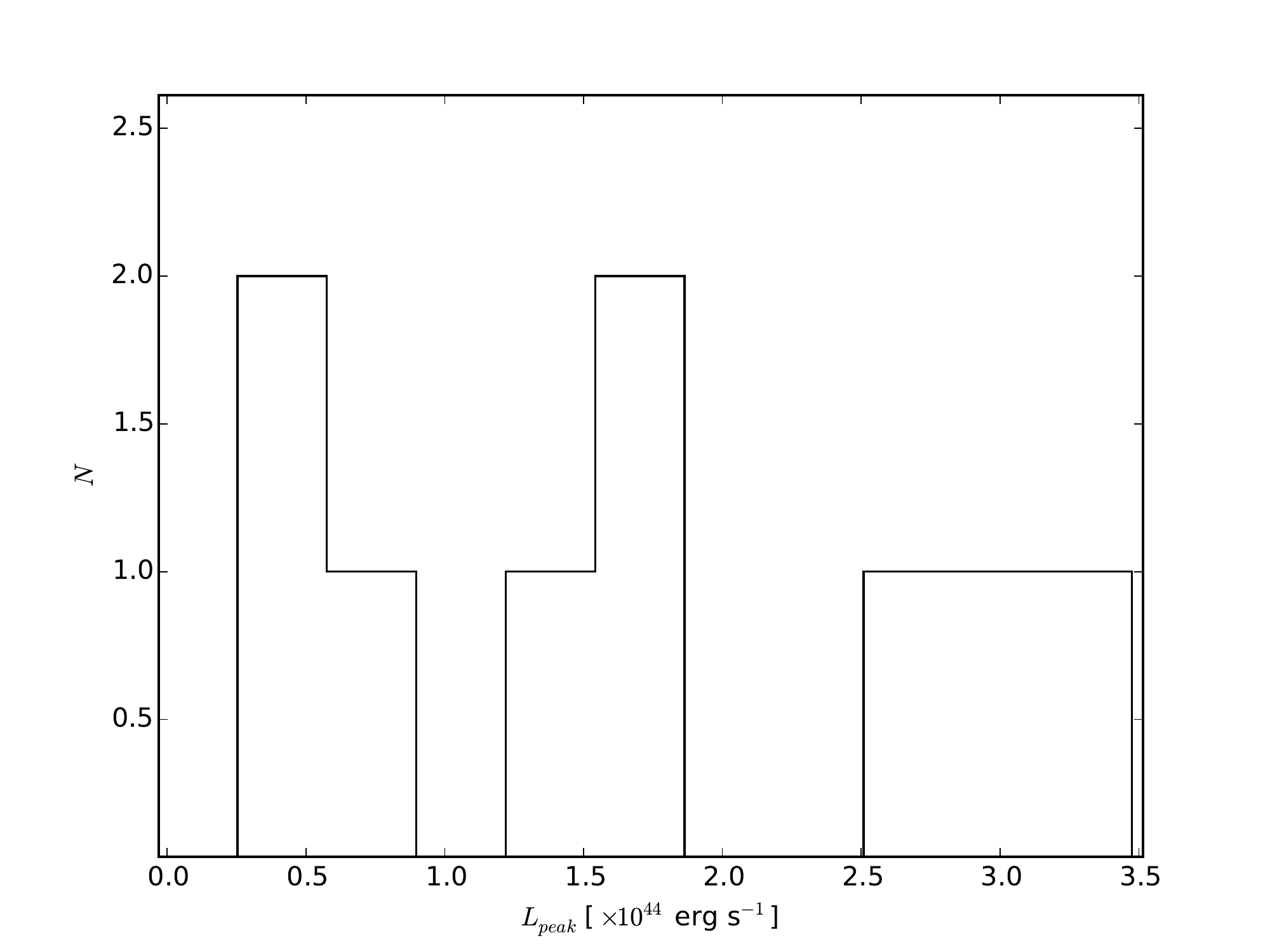}
\caption{Histogram of flare's peak luminosity in the 5-days binned BAT light curves in the TDE candidates sample.}
\label{fig:lpeak}
\end{figure}

The peak luminosities of the flares were estimated using the peak fluxes derived from the 5-day binned BAT count-rate lightcurves, converted to physical fluxes using the Crab as a template, as well as the distances to the host galaxies. The histogram of these luminosities, presented in Fig.~\ref{fig:lpeak}, spans values $10^{43-44}\, {\rm erg}\,{\rm s}^{-1}$, as expected for black holes of $10^{6-8}$ M$_\odot$.

The duration of the flares observed by BAT are between 40 and 110 days and their energy is in the range $10^{50-51}$ ergs. These are of course lower limits since TDEs are expected to last longer, below the BAT sensitivity. These durations are approximate and have a tendency to increase with the maximum flux of the flares, which qualitatively match the expectations.

During the six-month exposure of the {\it Rosat} All-Sky Survey, \cite{2002AJ....124.1308D} detected two TDEs from galaxies of the GWGC (NGC 5905 and IC 3599). 
Our hard X-ray TDE rate is, therefore, comparable to what was derived at soft X-rays. A more accurate comparison, taking into account the instrumental 
sensitivity limits and the GWGC sample is required. To derive the TDE distribution expected in a sample like the GWGC, we need to estimate the black hole masses in the host galaxies, the mass function of the disrupted objects, and the distribution of impact parameters.

A black-hole mass was estimated for each galaxy of the GWGC. 
We decided to use the total galaxy luminosity as a number of them lack morphological information and to account for irregular galaxies. The absolute B-band luminosities were converted to black hole masses following \cite{2013ApJ...764..184M}.
For the distribution of the disrupted objects masses $M_{*}$, we used the Kroupa initial mass function \citep[as parametrized in Equation 20 of][]{2014arXiv1410.7772S}.
The impact parameter $r_{p}$ distributions were derived for each black hole and disrupted object mass assuming isotropic encounters.

We defined a grid of parameter space with 5<log $M_{\rm BH}$<10 (step of 0.1), -1.1<log $M_{*}$ <1 (step of 0.01), and inverse of impact parameter -0.2<log $(\beta=r_t/r_p)$<0.6 (step of 0.01). The grid was limited in the region where the impact parameter is larger than the black hole gravitational radius.
On every point of that grid, we computed the peak accretion rate of a TDE, using equation A1 from \cite{2013ApJ...767...25G}, and converted it to a peak 
bolometric luminosity, using the radiative efficiency given by Fig. 1 in \cite{1999ApJ...514..180U}, and finally to a hard X-ray peak luminosity assuming 
that 4\% of the total emission is released at hard X-rays, as in active galactic nuclei \citep{2004ASSL..308..187R}.

Each point of the grid was weighted by the Kroupa IMF and by the probability of the impact parameter. The peak luminosity was then summed in the grid, using the black hole mass distribution derived from the GWGC as a function of the distance to finally obtain the peak flux distribution. 
This distribution was renormalised such that the number of events predicted above $2\times 10^{-10}\,{\rm erg}\,{\rm s}^{-1}\,{\rm cm}^{-2}$ (our BAT sensitivity limit) corresponds to the observed hard X-ray TDE rate.

\begin{figure}
\centering
\includegraphics[width=0.49\textwidth]{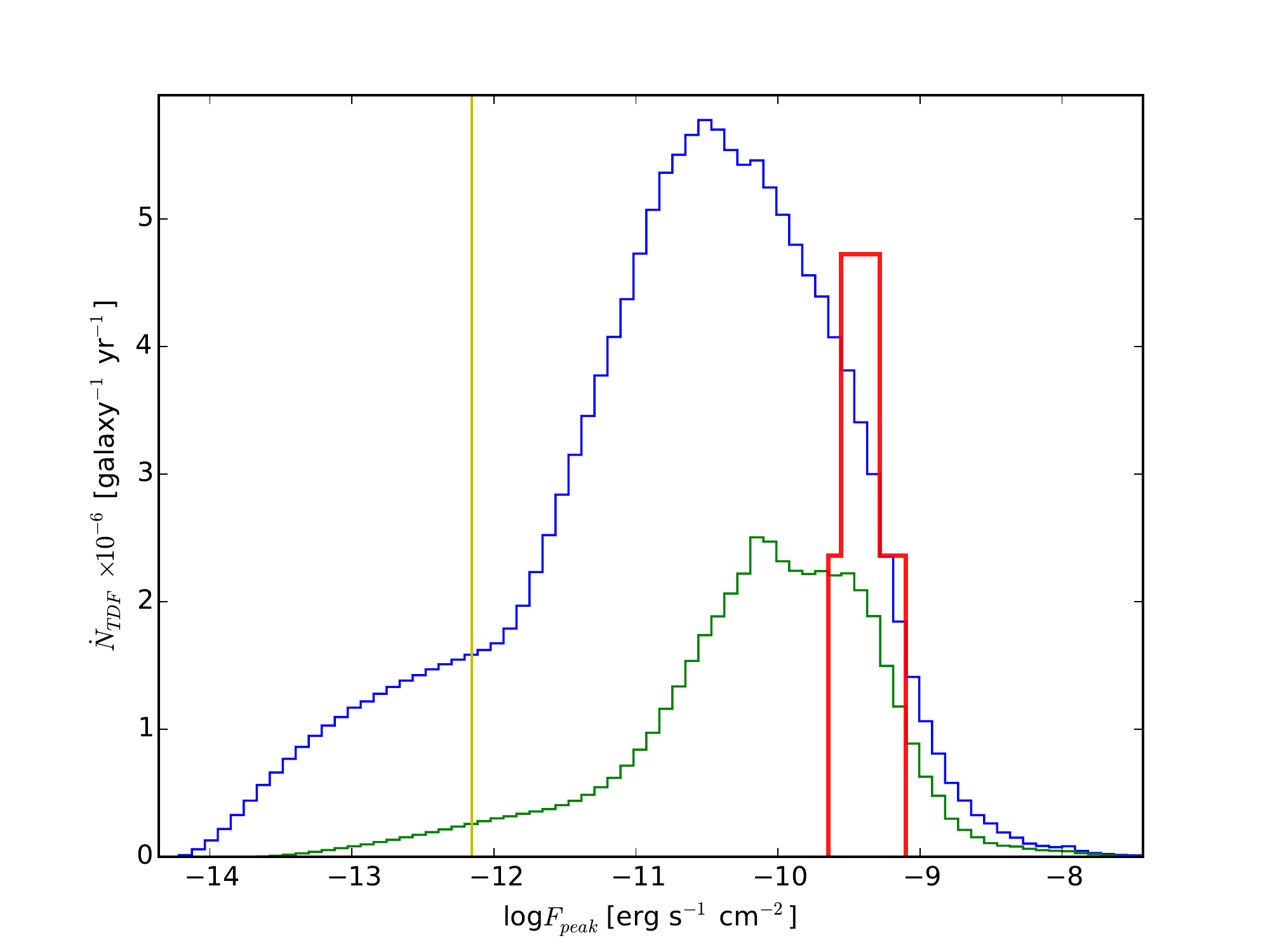}
\caption{Expected distribution of peak fluxes for TDE in the GWGC for all black holes (blue) and only from those with masses in the range $10^{6-7}$ M$_\odot$ (green). The red histogram shows the TDE candidates detected by us. The vertical line shows the expected {\it eRosita} flare sensitivity at hard X-rays.}
\label{fig:flux}
\end{figure}

The resulting distribution is shown in the blue histogram in Fig.~\ref{fig:flux}, which shows two components corresponding to the change of equation of state between low and high mass stars. The green histogram shows the contribution of black hole masses of $10^{6-7}$ M$_\odot$. The red histogram is derived from the observed sample. The shape of the red histogram does not reproduce the high flux tail of the blue histogram but we should remember that above a flux of $10^{-9}\,{\rm erg}\,{\rm s}^{-1}\,{\rm cm}^{-2}$, the blue histogram predicts one TDE every five years for the GWGC, which is consistent with our non-detection. The blue distribution is also based on several assumptions (like black-hole mass and IMF) that may not be representative for the GWGC sample, for instance three of our nine TDE candidates occurred in irregular galaxies.

\begin{figure}
\centering
\includegraphics[width=0.49\textwidth]{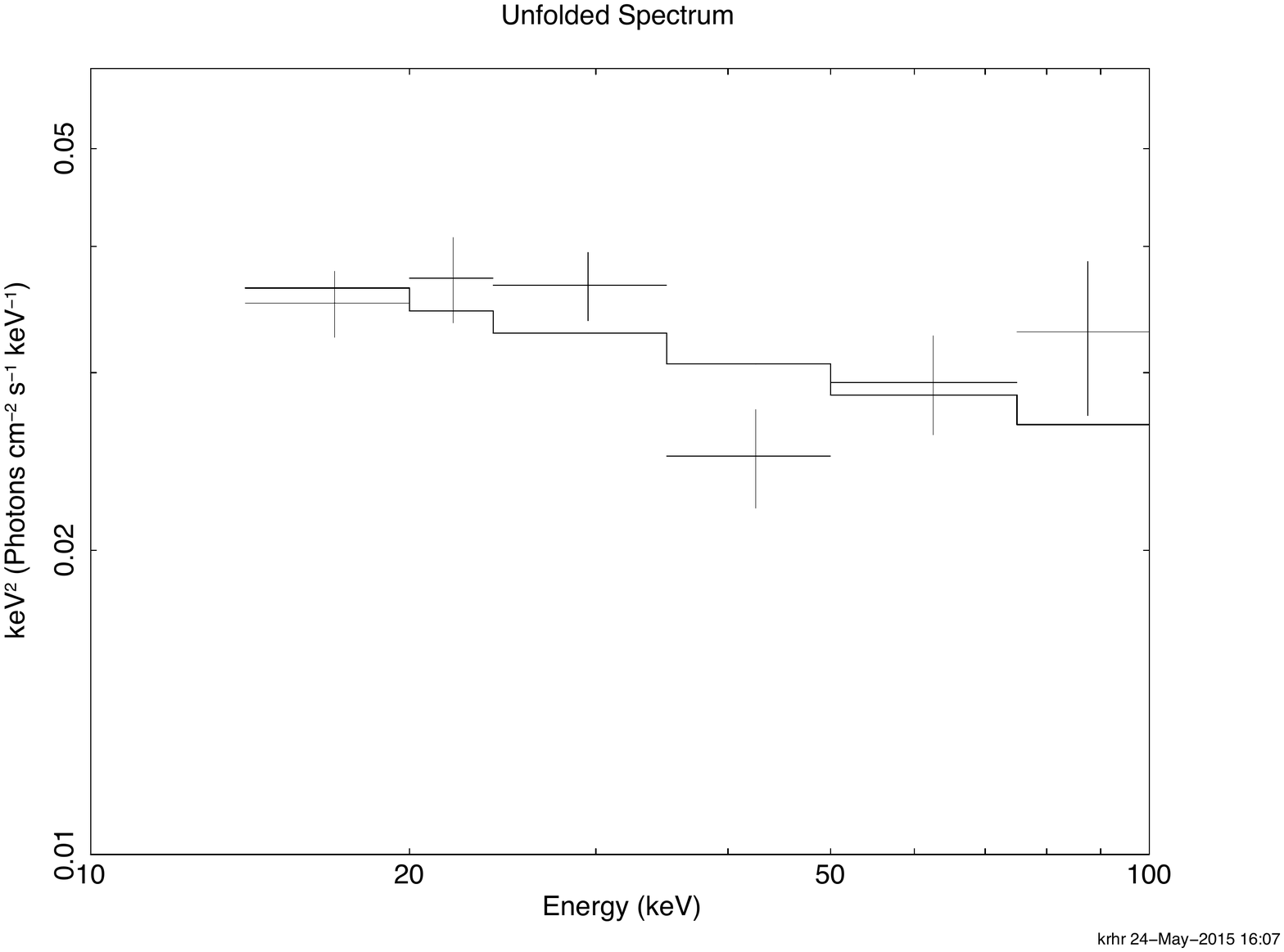}
\caption{Average spectrum of the nine BAT TDEs.}
\label{fig:spec}
\end{figure}

Some interesting numbers can be derived from Fig. \ref{fig:flux} and be compared with previous TDE rate estimates.
The TDE rate, which is derived from the blue histogram above a flux of $2.5\times 10^{-13}\,{\rm erg}\,{\rm s}^{-1}\,{\rm cm}^{-2}$ is $1.3 \times 10^{-4}\,{\rm galaxy}^{-1}\,{\rm yr}^{-1}$,
matches well the rate of $2.3 \times 10^{-4}\,{\rm galaxy}^{-1}\,{\rm yr}^{-1}$ \citep{2008A&A...489..543E} derived from the {\it XMM-Newton} slew survey (and optical flares as well).
The TDE rate, for black holes in the range $10^{6-7}$ M$_\odot$, which was obtained by integrating the green histogram, is $4.9 \times 10^{-5}\,{\rm galaxy}^{-1}\,{\rm yr}^{-1}$, which  also matches  the expectations  from the simulations of \cite{2011MNRAS.418.1308B} well.
The TDE rate derived from the blue histogram above a flux of $10^{-12}\,{\rm erg}\,{\rm s}^{-1}\,{\rm cm}^{-2}$ (the {\it Rosat} flux limit) is $1.2 \times 10^{-4}\,{\rm galaxy}^{-1}\,{\rm yr}^{-1}$, which is significantly higher than the observed rate of $3 \times 10^{-5}\,{\rm galaxy}^{-1}\,{\rm yr}^{-1}$ \citep{2014MNRAS.444.1041K}. As the latter is also in tension with the {\it XMM-Newton} observations, it is difficult to conclude if this has a specific meaning.

Starting in 2016, {\it eRosita} \citep{2012arXiv1209.3114M} will scan the X-ray sky regularly, observing any sky position every six months or less. Taking into account that one TDE flare could be detected for three months, we expect seven TDE per year to be detected at hard X-rays for GWGC galaxies and more from higher redshift sources. Most TDEs that originate in the GWGC will be detected at soft X-rays, which will give an observational access to most of the histogram featured in Fig. \ref{fig:flux}.

We have built the average spectrum of our nine TDE candidates by extracting the spectrum of each individual flare and averaging them.
The average spectrum, unfolded using a powerlaw model, is shown in Fig. \ref{fig:spec}. The average spectrum is well fitted by a power law 
with an index of $2.18\pm0.08$ and an average flux of $7.8\times 10^{-11}\,{\rm erg}\,{\rm s}^{-1}\,{\rm cm}^{-2}$ (20-100 keV). The power-law slope 
is similar to that observed in active galactic nuclei.

\section{Conclusions}

We have detected nine TDE candidates at hard X-rays during eight years in a sample of $\sim 53\,000$ galaxies closer to 100 Mpc. These events are not beamed as their peak luminosities ($3-34\times 10^{43}\,{\rm erg}\,{\rm s}^{-1}$) are not significantly higher than the Eddington limits ($10^{43.8-46}\,{\rm erg}\,{\rm s}^{-1}$) derived from the corresponding black hole masses.

Assuming that 4\% of the TDE bolometric luminosity is emitted at hard X-rays, the observed hard X-ray fluxes and rate are consistent with the rates observed by {\it XMM-Newton}  at soft X-rays or in the optical from SDSS observations. It is also consistent with expectations from simulations.

We conclude that hard X-ray emission should be ubiquitous in unbeamed TDEs. The average hard X-ray spectral shape of our TDE candidates is, in addition, very similar to that usually observed in active galactic nuclei, which points towards inverse Compton processes and electrons that are accelerated in shocks in the accretion flow.
We also discovered a long flare of RX J1709.5-2639 and two new persistent active galactic nuclei candidates.

\bibliographystyle{aa} 
\bibliography{tdeflares} 



\begin{figure*}
\includegraphics[angle=90,width=0.19\textwidth]{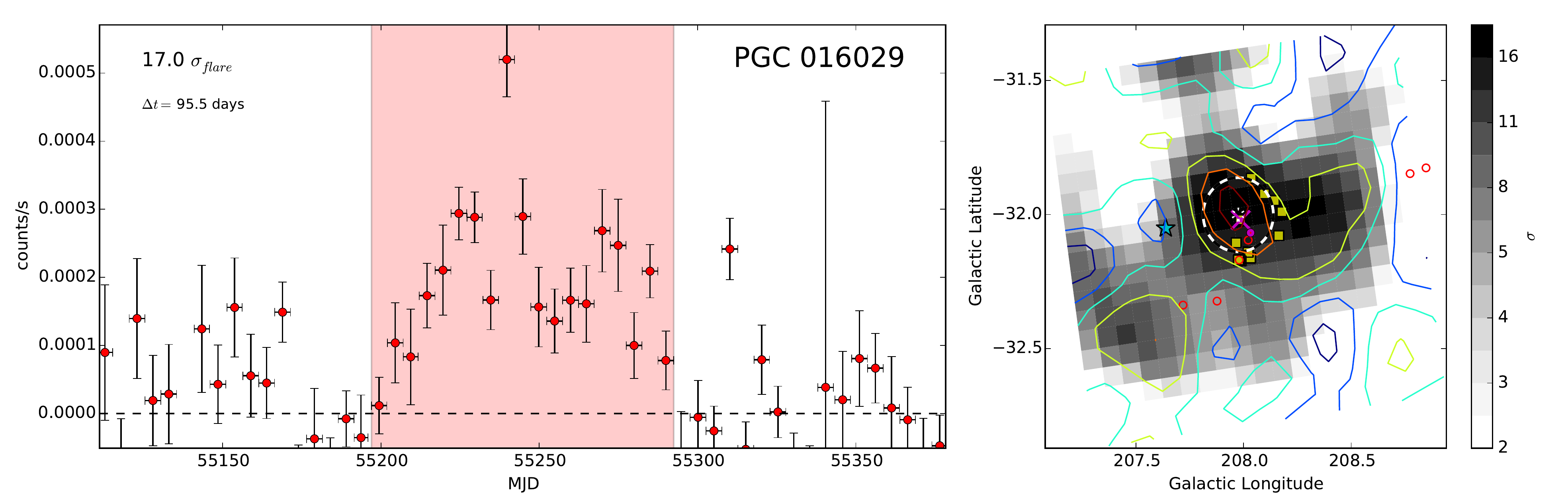}
\includegraphics[angle=90,width=0.19\textwidth]{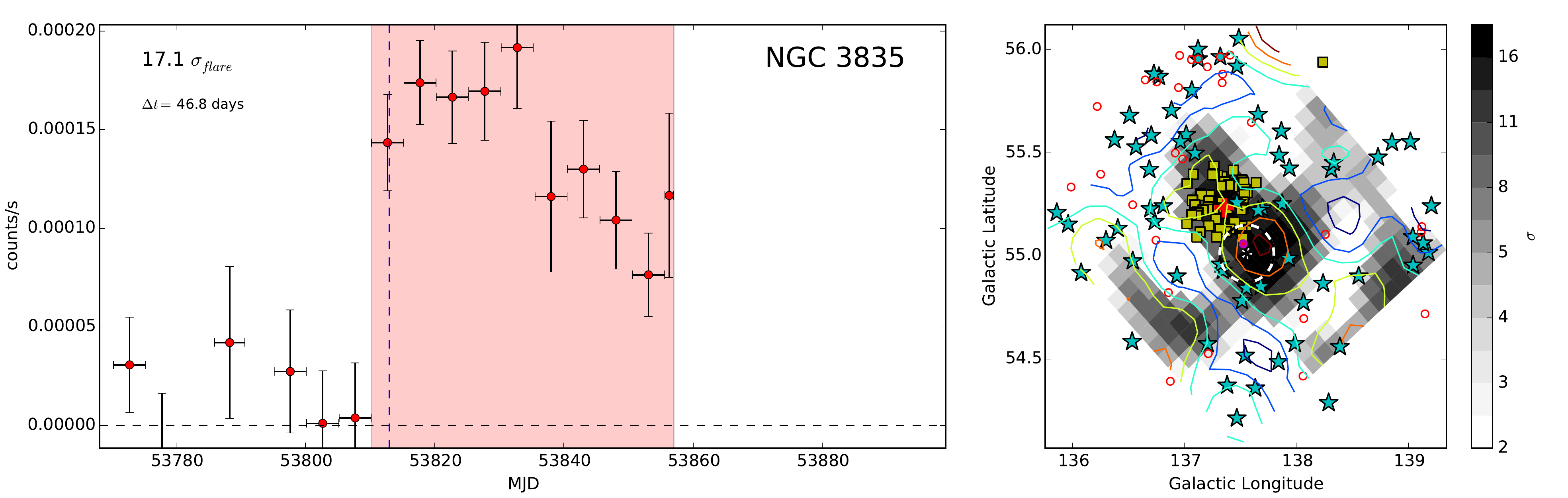}
\includegraphics[angle=90,width=0.19\textwidth]{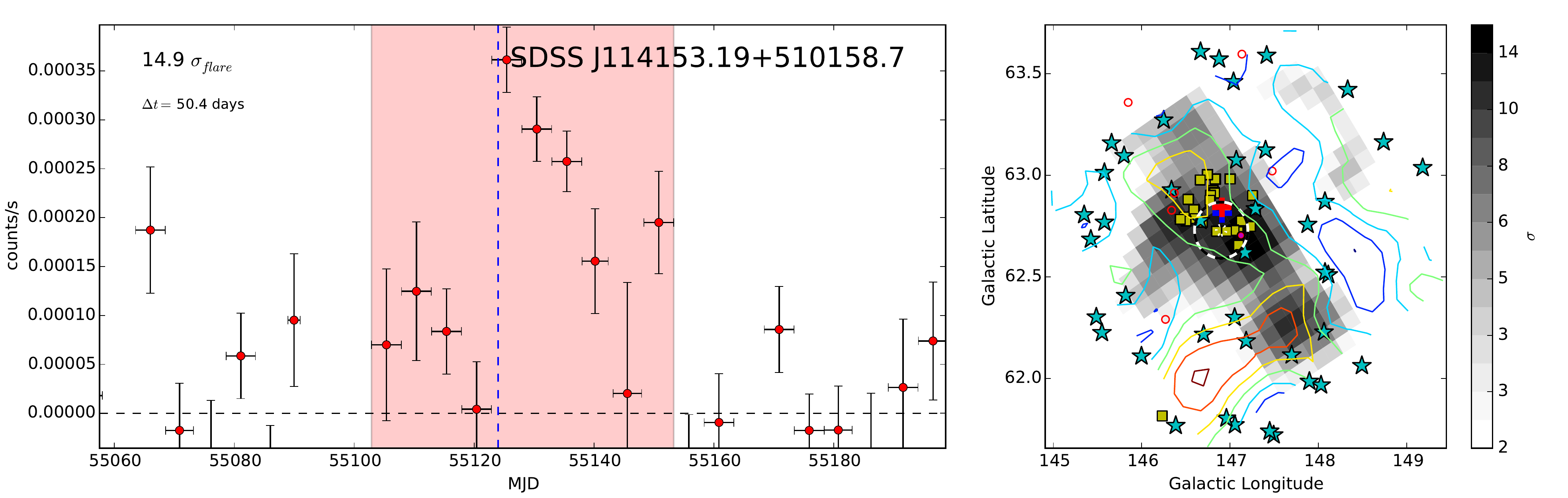}
\includegraphics[angle=90,width=0.19\textwidth]{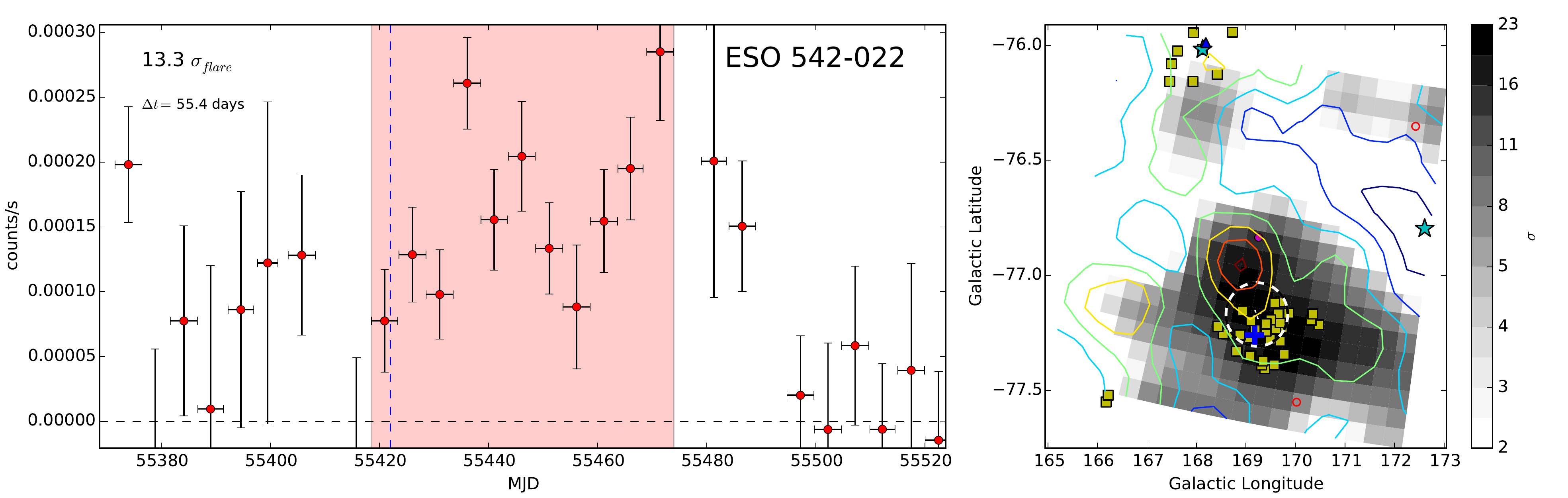}
\includegraphics[angle=90,width=0.19\textwidth]{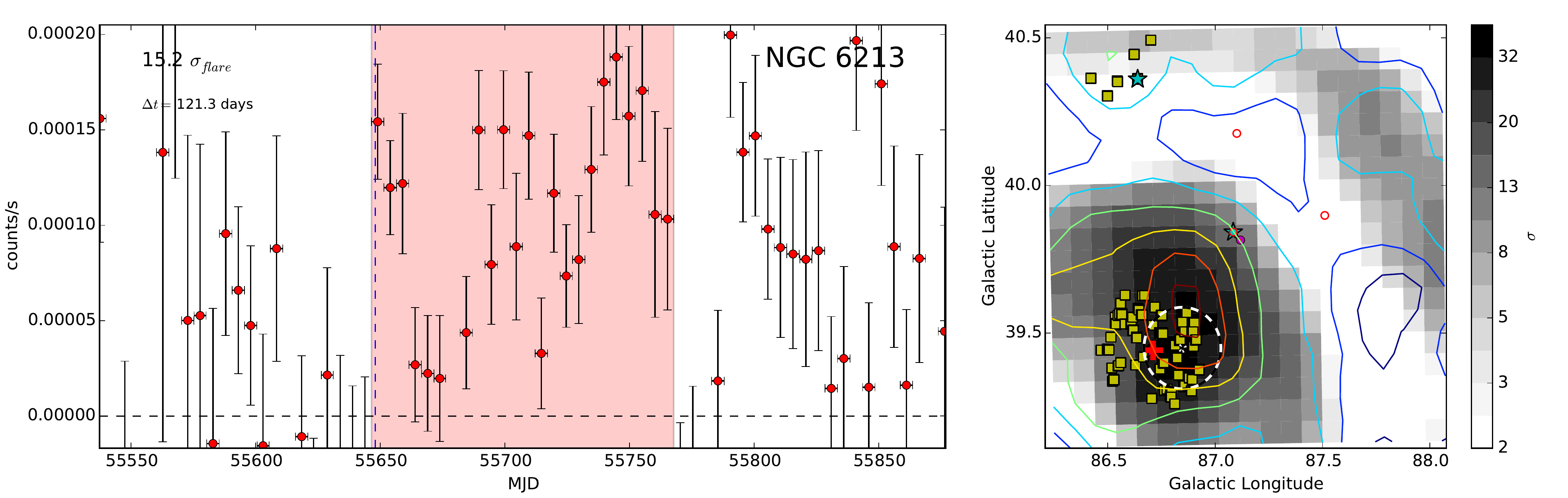}

\includegraphics[angle=90,width=0.19\textwidth]{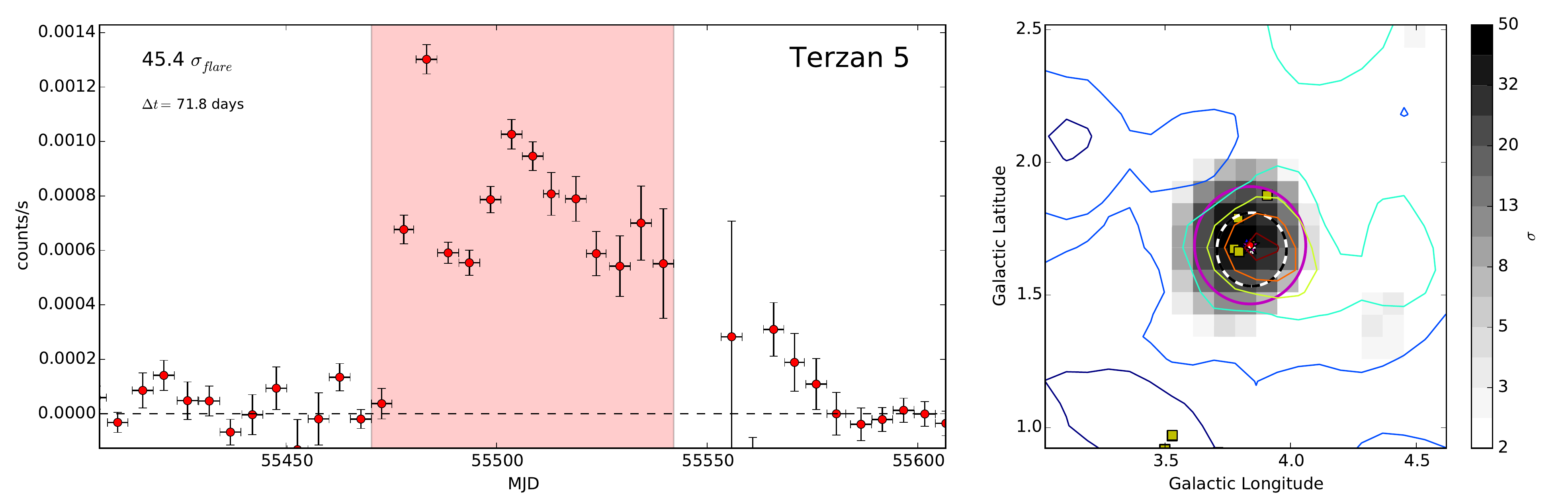}
\includegraphics[angle=90,width=0.19\textwidth]{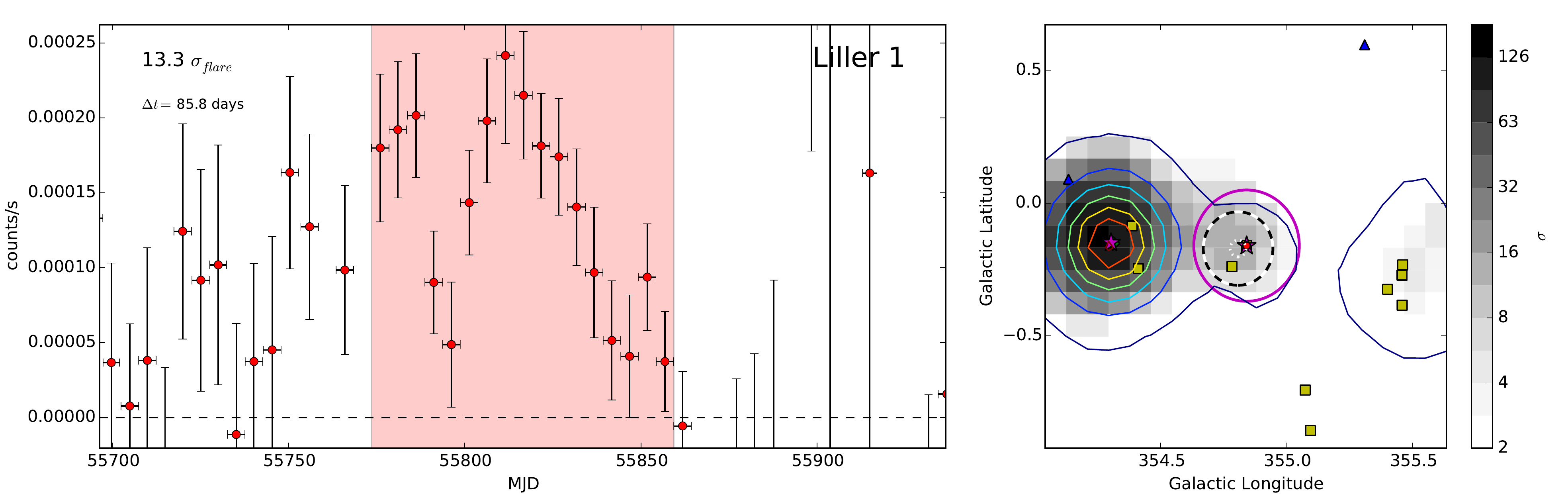}
\includegraphics[angle=90,width=0.19\textwidth]{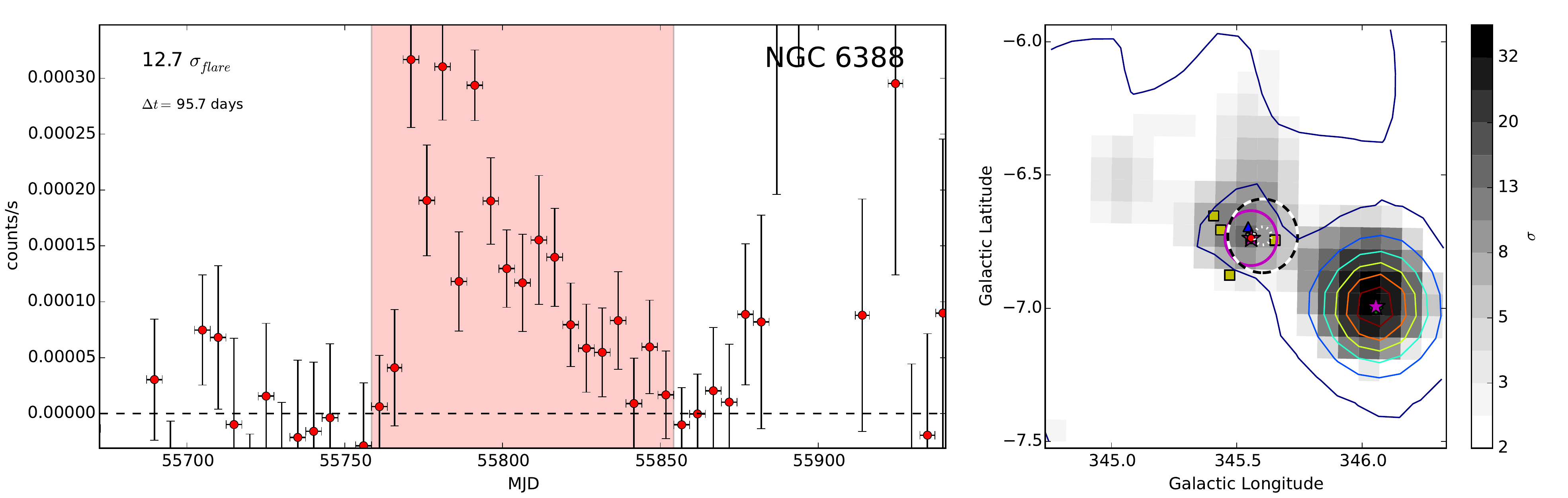}
\includegraphics[angle=90,width=0.19\textwidth]{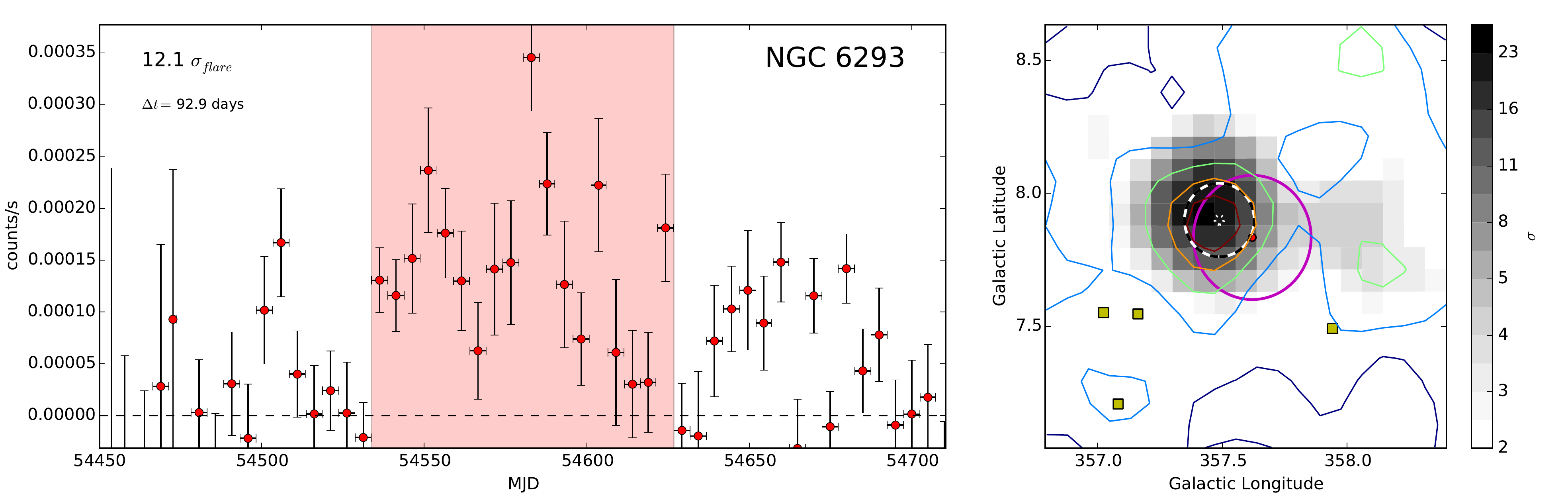}
\includegraphics[angle=90,width=0.19\textwidth]{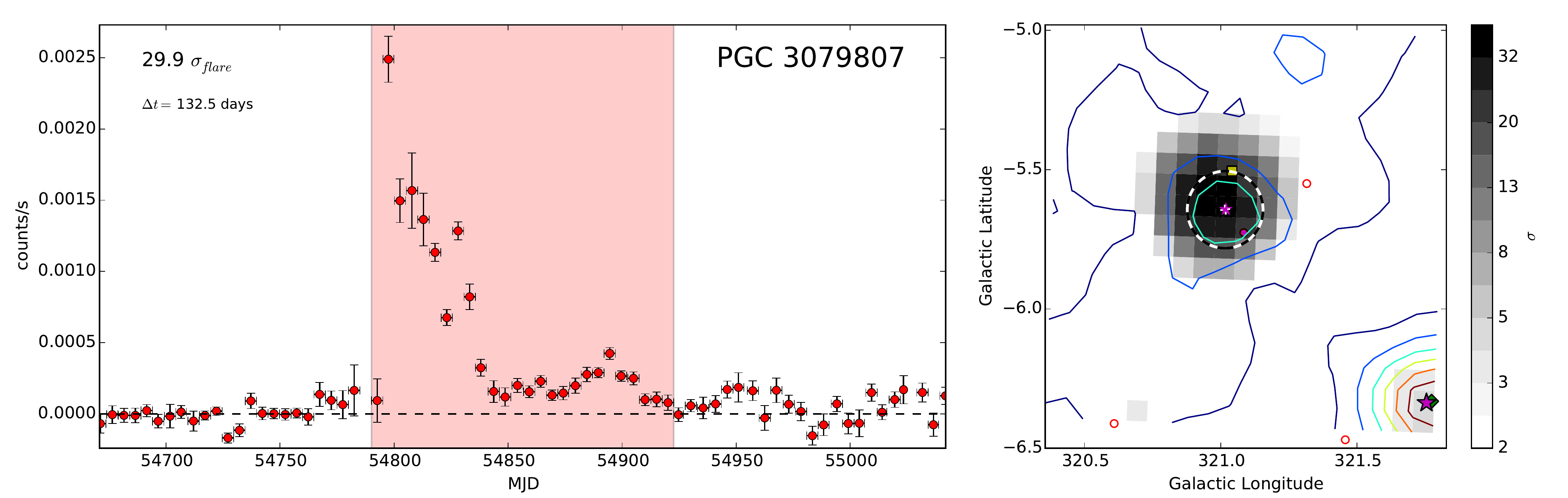}
\caption{{\bf Left panels:} the light curves derived at the position of the flares maximum, with red individual points with error bars, while the detected flare is shaded with red. If a GRB was detected nearby, a target galaxy vertical dashed line shows the time of the GRB. {\bf Right panels:} significance images built over the flare period in logarithmic greyscale (the lower limit is fixed at 2 $\sigma$). The associated GWGC galaxies are marked with red-filled circles, while other galaxies from the catalogue are shown with  empty red circles. Cyan stars mark AGNs, yellow squares are XRT point sources, green diamonds show BAT galactic sources, blue triangles are {\it Fermi} sources, blue and red + show GRBs from {\it Fermi} and BAT, cyan x shows supernovae candidates, magenta x mark classical novae detections, red x code bright BAT sources, and magenta stars present BAT transients. The big black empty circle represents PSF fit with its radius equal sigma of the Gaussian (8.2 arcsec) with inner circle with radius equal to the error of the PSF. Magenta circles shows galaxy/globular cluster size if it was given in the catalogue. Overplotted contours are significance contours of the 8-year images.}
\label{fig:obj}
\end{figure*}
\begin{figure*}
\includegraphics[angle=90,width=0.19\textwidth]{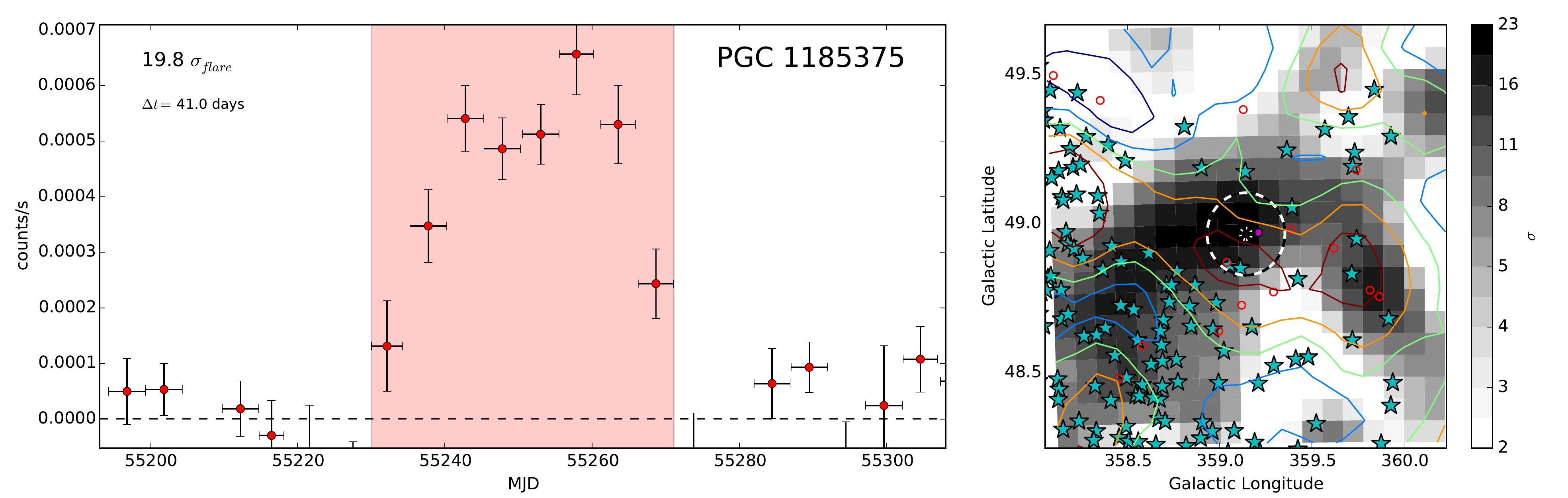}
\includegraphics[angle=90,width=0.19\textwidth]{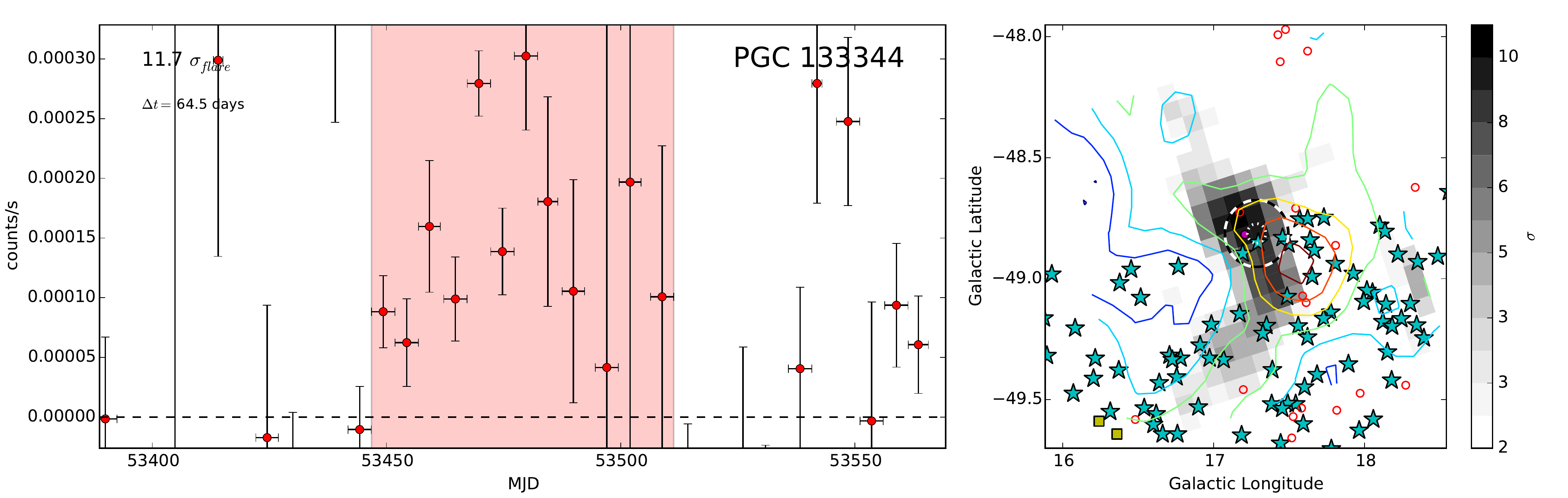}
\includegraphics[angle=90,width=0.19\textwidth]{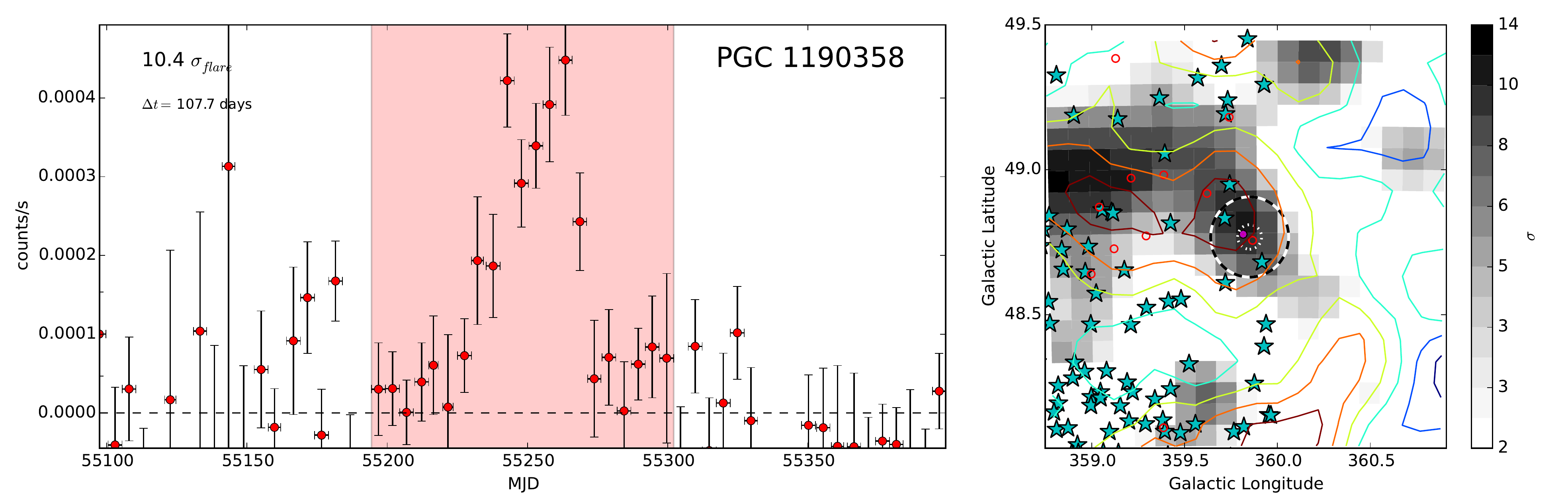}
\includegraphics[angle=90,width=0.19\textwidth]{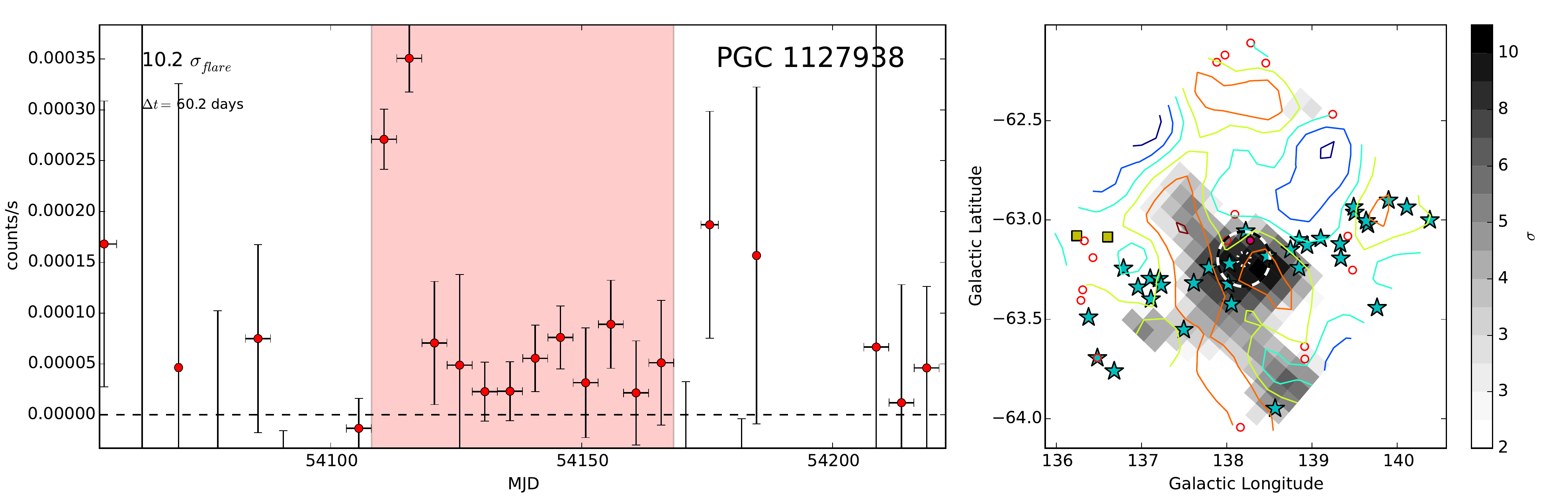}
\includegraphics[angle=90,width=0.19\textwidth]{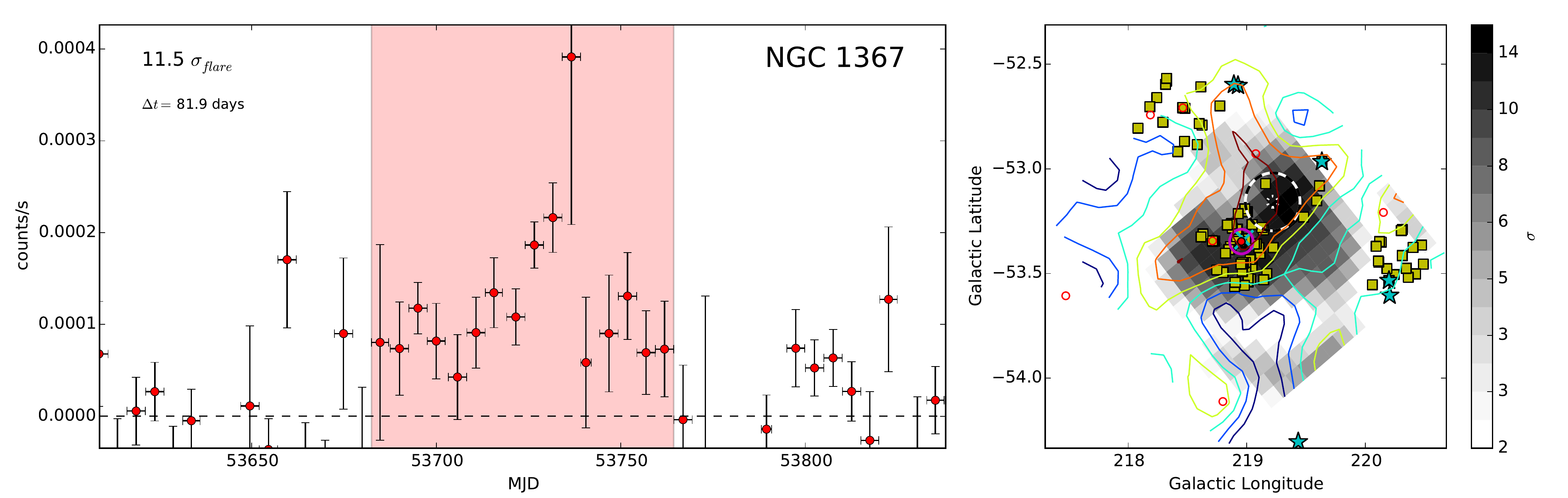}

\includegraphics[angle=90,width=0.19\textwidth]{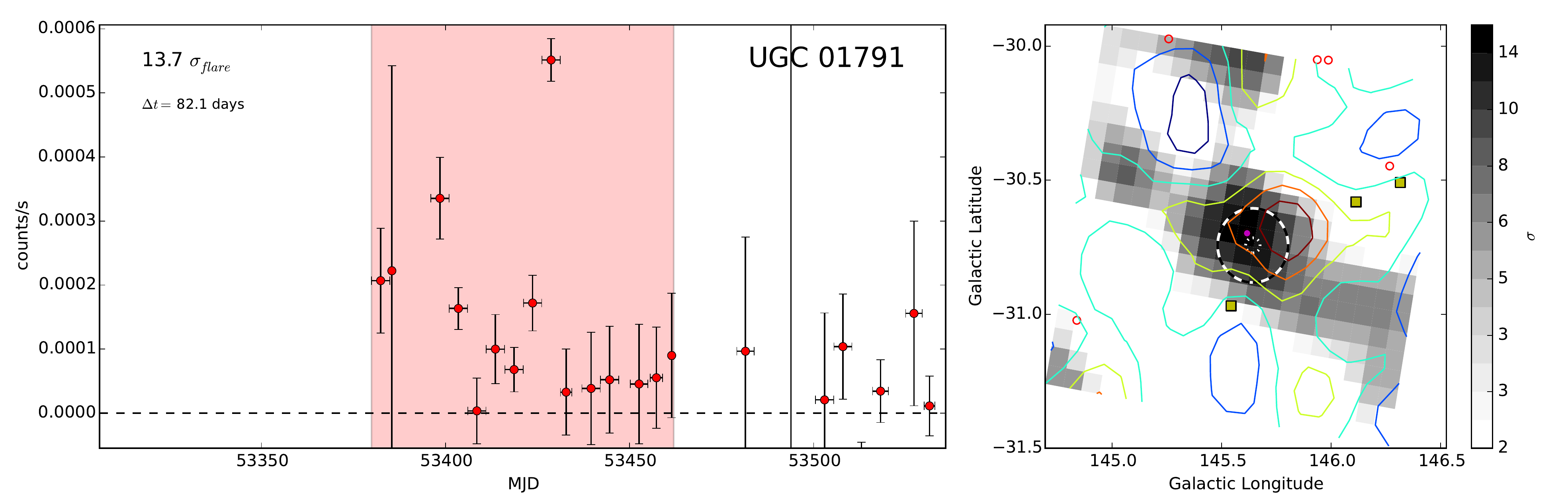}
\includegraphics[angle=90,width=0.19\textwidth]{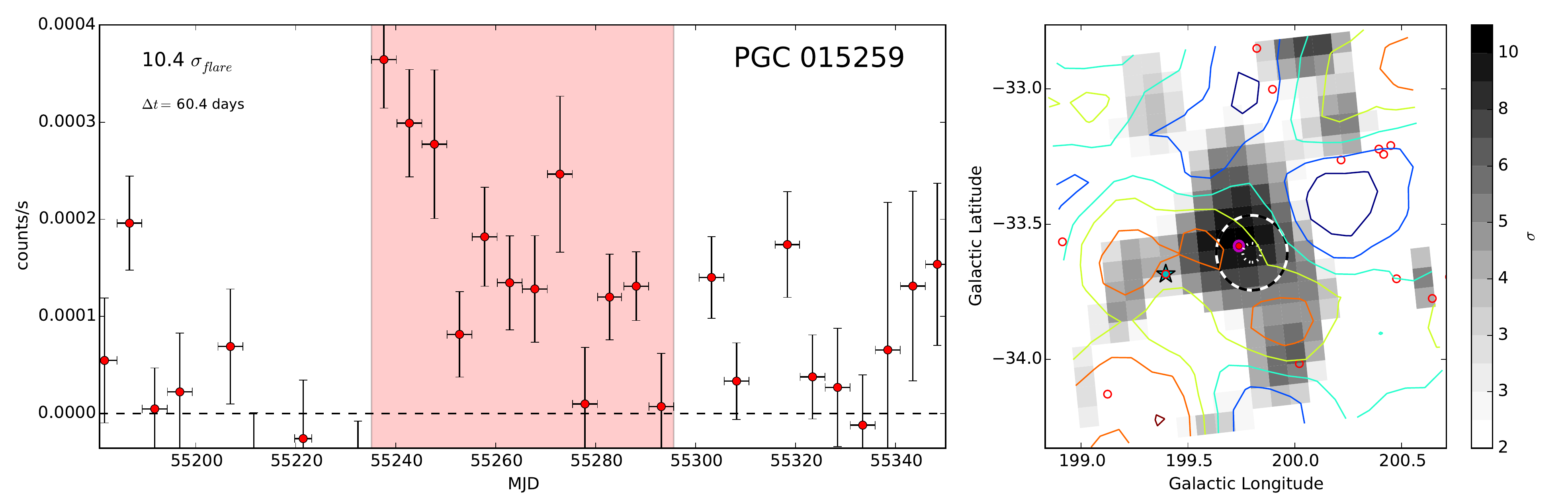}
\includegraphics[angle=90,width=0.19\textwidth]{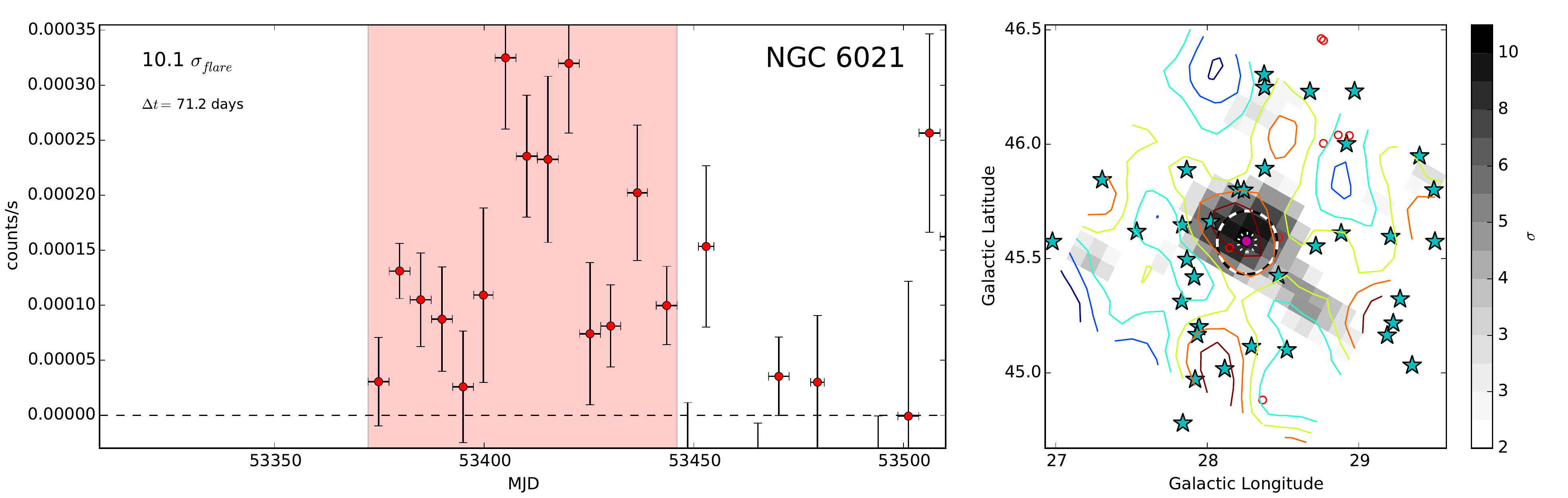}
\includegraphics[angle=90,width=0.19\textwidth]{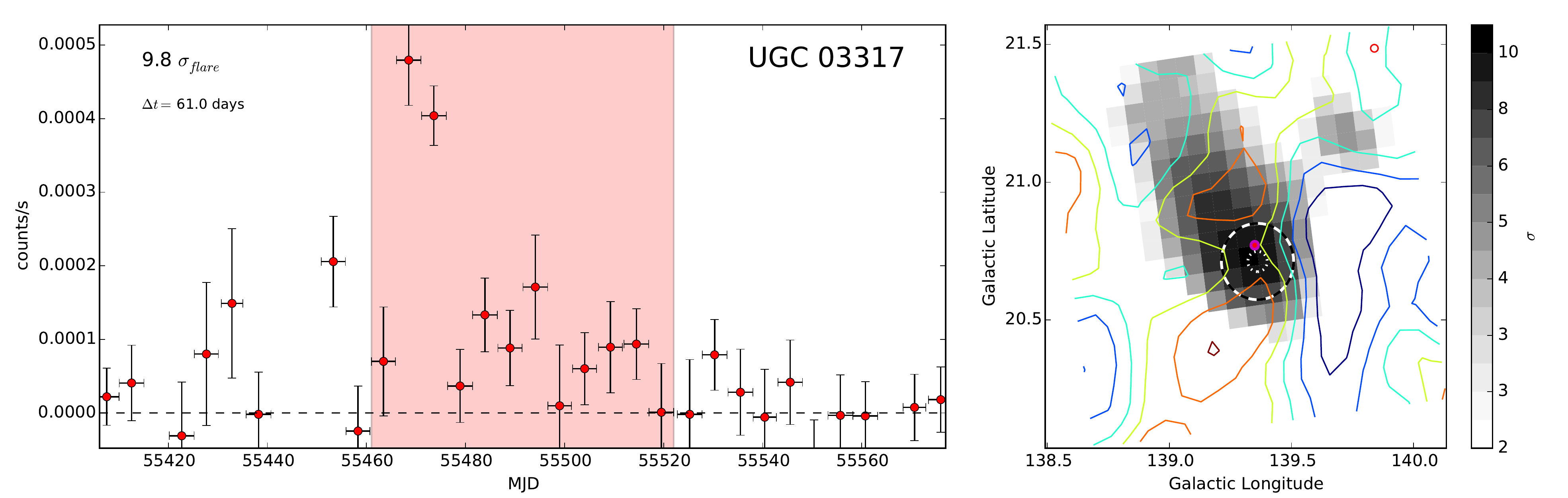}
\includegraphics[angle=90,width=0.19\textwidth]{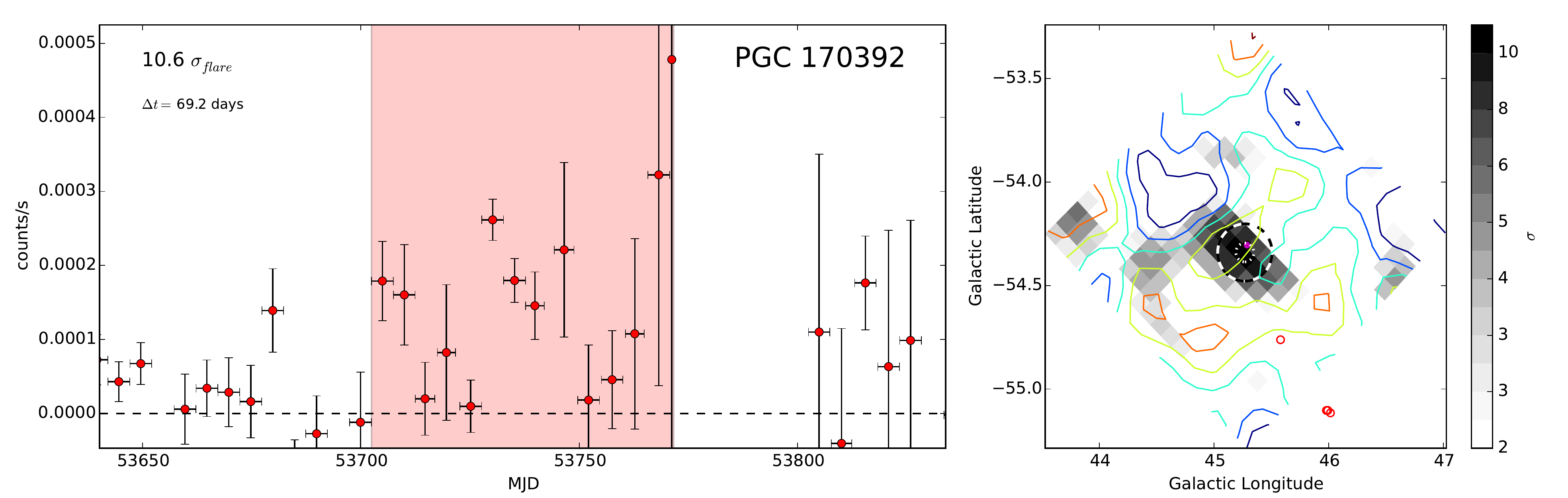}
Fig.~\ref{fig:obj}. (continued)
\end{figure*}
\begin{figure*}
\includegraphics[angle=90,width=0.19\textwidth]{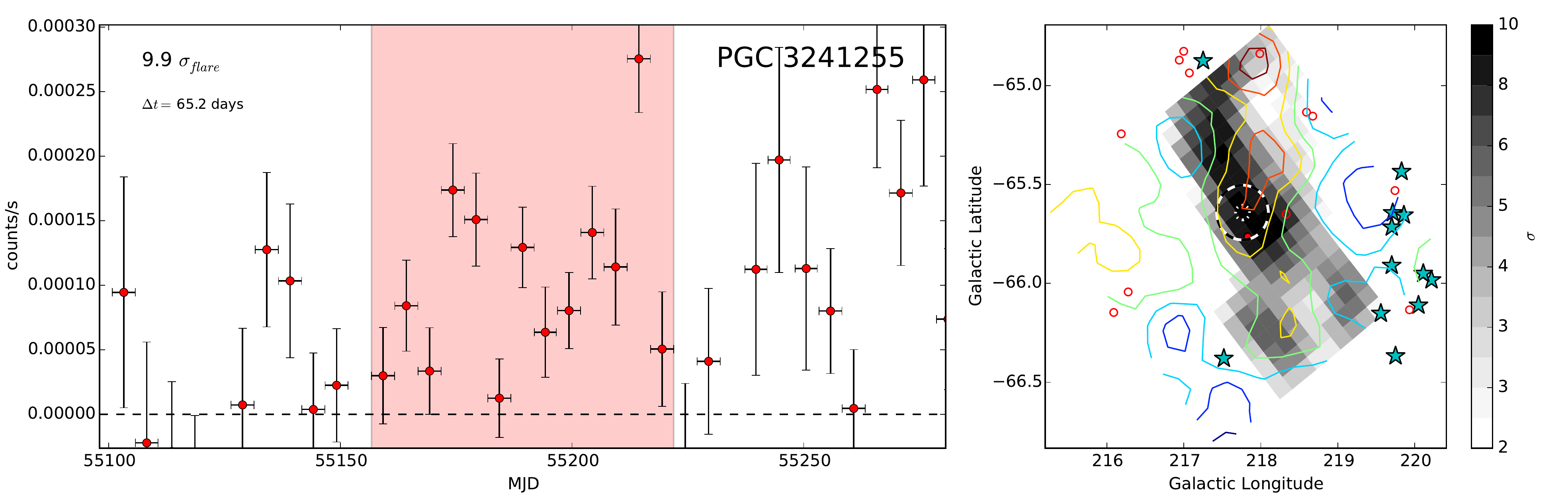}

\includegraphics[angle=90,width=0.19\textwidth]{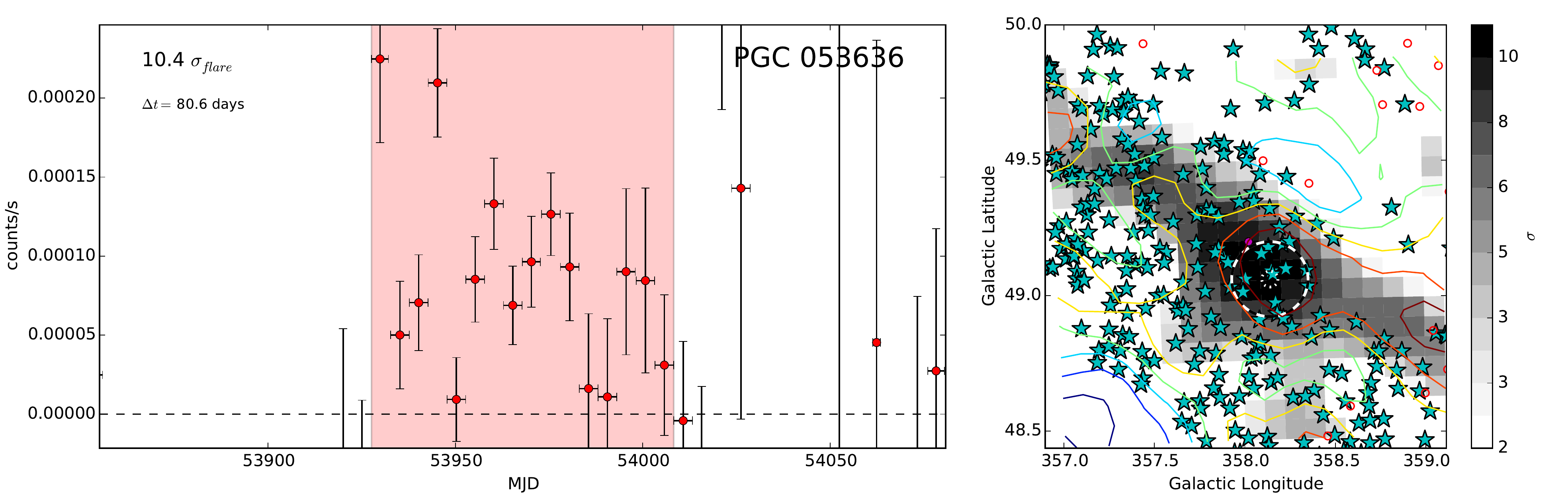}
\includegraphics[angle=90,width=0.19\textwidth]{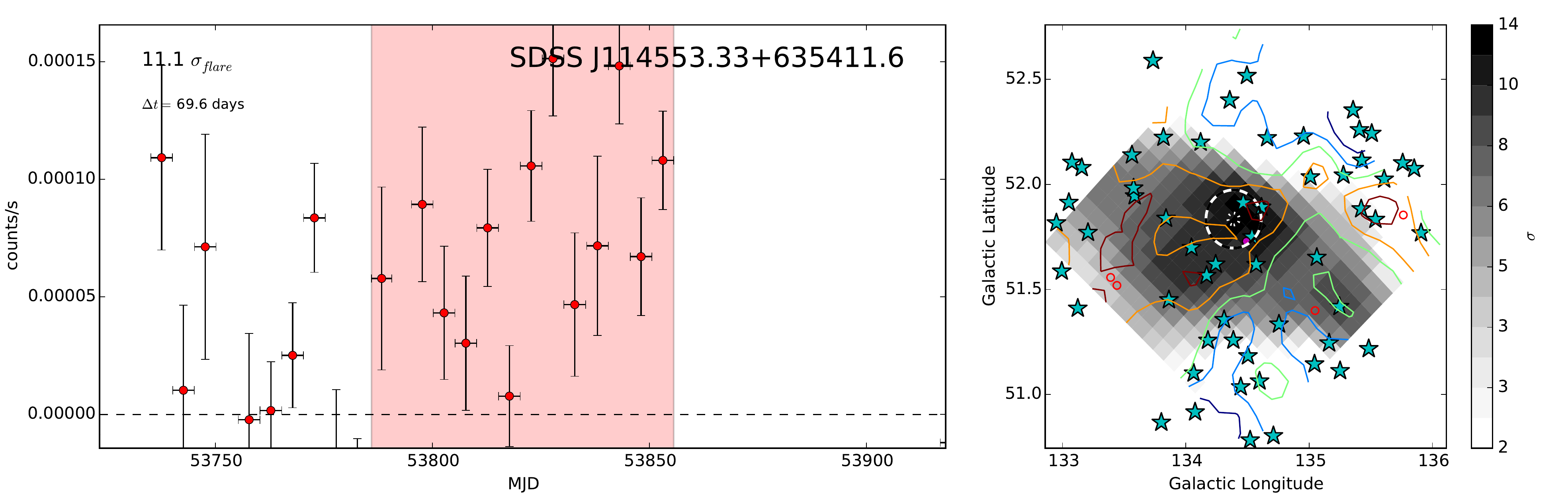}
\includegraphics[angle=90,width=0.19\textwidth]{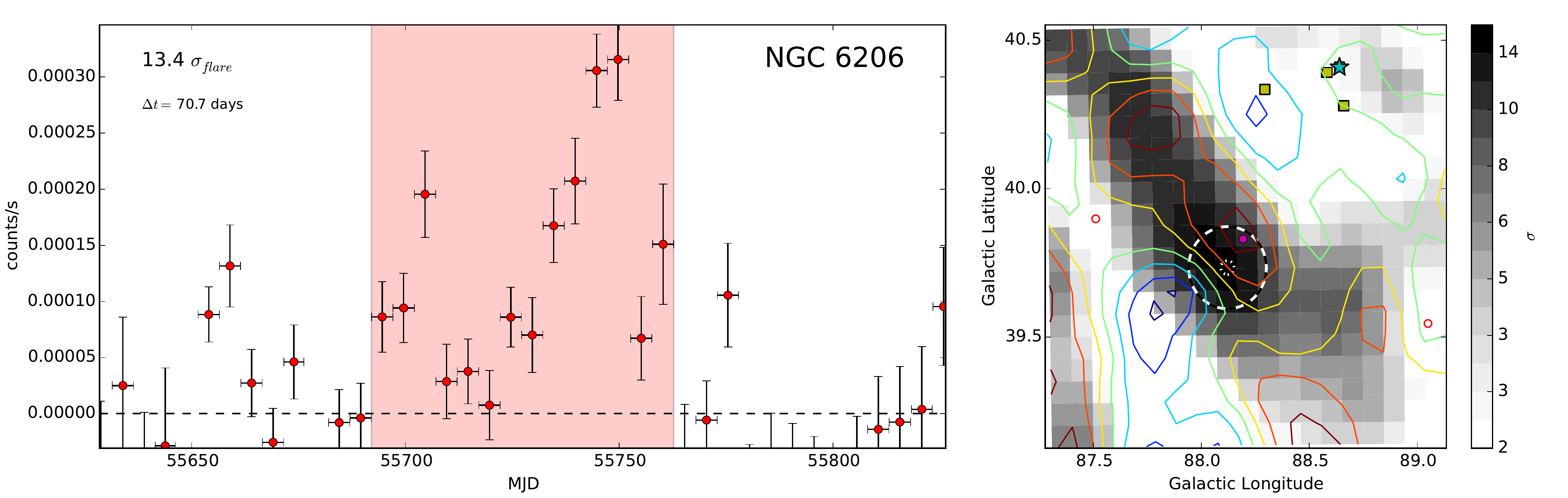}
\includegraphics[angle=90,width=0.19\textwidth]{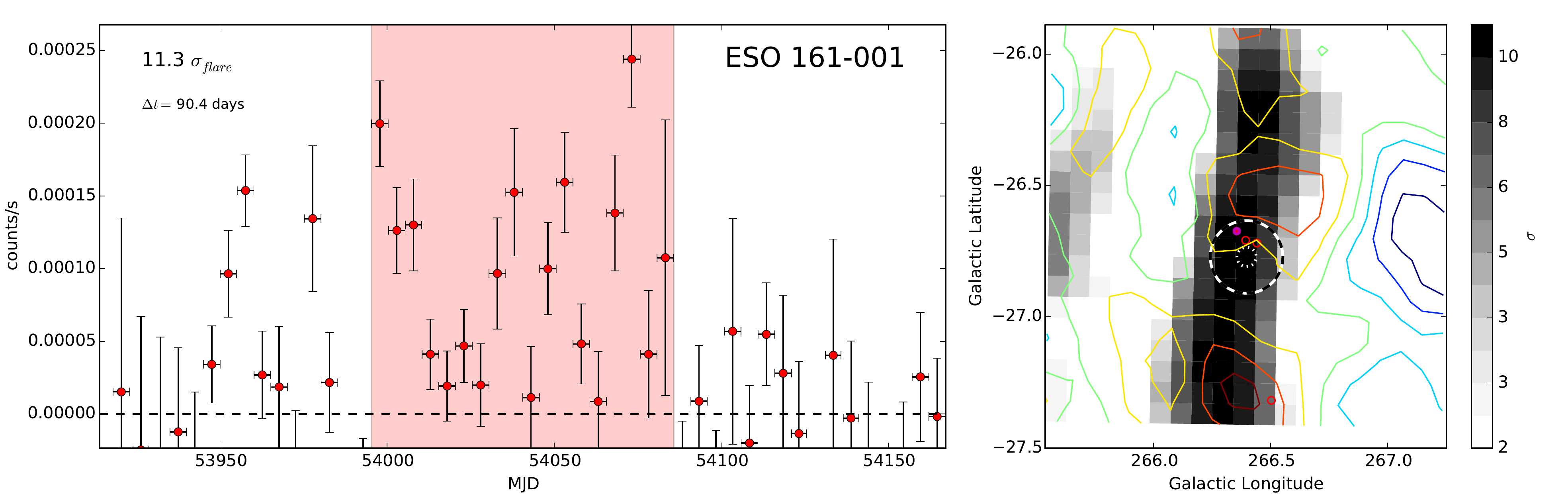}
\includegraphics[angle=90,width=0.19\textwidth]{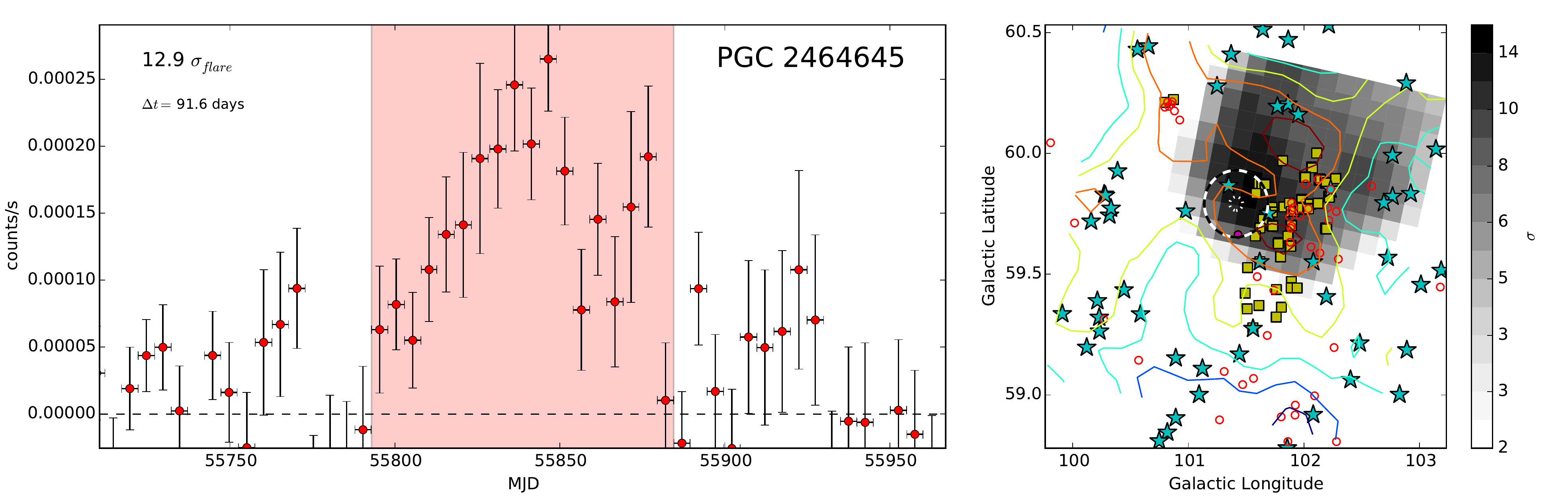}
Fig.~\ref{fig:obj}. (continued)
\end{figure*}

\end{document}